\documentclass[manuscript,screen,sigconf,nonacm]{acmart}

\usepackage{hyperref}
\usepackage{url}
\usepackage{graphicx}
\usepackage{cleveref}
\usepackage{subcaption}
\usepackage[absolute]{textpos}
\AtBeginDocument{
  }

\begin{document}

\begin{textblock*}{\textwidth}(0.75in, 0.5in)
  \noindent A shortened version of this paper appears in the \textit{Proceedings of the 4th ACM Conference on Equity and Access in Algorithms, Mechanisms, and Optimization (EAAMO ‘24)}, October 2024. This is the extended version.
\end{textblock*}

\title{Who's in and who's out? A case study of multimodal CLIP-filtering in DataComp}

\author{Rachel Hong}
\email{hongrach@cs.washington.edu}
\affiliation{
  \institution{University of Washington}
  \city{Seattle}
  \state{Washington}
  \country{USA}
}

\author{William Agnew}
\affiliation{
  \institution{Carnegie Mellon University}
  \city{Pittsburgh}
  \state{Pennsylvania}
  \country{USA}
}

\author{Tadayoshi Kohno}
\affiliation{
  \institution{University of Washington}
  \city{Seattle}
  \state{Washington}
  \country{USA}
}

\author{Jamie Morgenstern}
\affiliation{
  \institution{University of Washington}
  \city{Seattle}
  \state{Washington}
  \country{USA}
}

\renewcommand{\shortauthors}{Hong et al.}

\begin{abstract}
As training datasets become increasingly drawn from unstructured, uncontrolled environments such as the web, researchers and industry practitioners have increasingly relied upon data filtering techniques to ``filter out the noise'' of web-scraped data. While datasets have been widely shown to reflect the biases and values of their creators, in this paper we contribute to an emerging body of research that assesses the filters used to create these datasets. We show that image-text data filtering also has biases and is value-laden, encoding specific notions of what is counted as ``high-quality'' data. In our work, we audit a standard approach of image-text CLIP-filtering on the academic benchmark DataComp's CommonPool by analyzing discrepancies of filtering through various annotation techniques across multiple modalities of image, text, and website source. We find that data relating to several imputed demographic groups --- such as LGBTQ+ people, older women, and younger men --- are associated with higher rates of exclusion. We also find prevalence of \textit{Western bias}, where the CLIP filter is more likely to include data related to Western countries compared to that of non-Western countries. Moreover, we demonstrate cases of \textit{exclusion amplification}: not only are certain marginalized groups already underrepresented in the unfiltered data, but CLIP-filtering excludes data from these groups at higher rates. The data-filtering step in the machine learning pipeline can therefore exacerbate representation disparities already present in the data-gathering step, especially when existing filters are designed to optimize a specifically-chosen downstream performance metric like zero-shot image classification accuracy. Finally, we show that the NSFW filter fails to remove sexually-explicit content from CommonPool, and that CLIP-filtering includes several categories of copyrighted content at high rates. Our conclusions point to a need for fundamental changes in dataset creation and filtering practices.
\textcolor{red}{Content warning: This paper discusses societal stereotypes and sexually-explicit material that may be disturbing, distressing, and/or offensive to the reader.}
\end{abstract}



\received{15 April 2024}

\maketitle

\section{Introduction}
\label{intro}

Text and image datasets in machine learning have grown to the scale of billions \citep{bert, gpt2, laion5b} in order to build larger text and image generative models \citep{scale, stablediffusion}. To reach such magnitudes, datasets are being increasingly scraped from the web \citep{c4t5, datacomp}, through archives such as Common Crawl \citep{commoncrawl}. While data from these web dumps are expansive, they are not expertly curated and may reduce the efficacy of downstream models. Manually assessing every data point is infeasible, however, so this requires at-scale methods to remove data that curators deem undesirable.

\begin{figure*}
    \begin{center}
    \includegraphics[width=0.9\linewidth]{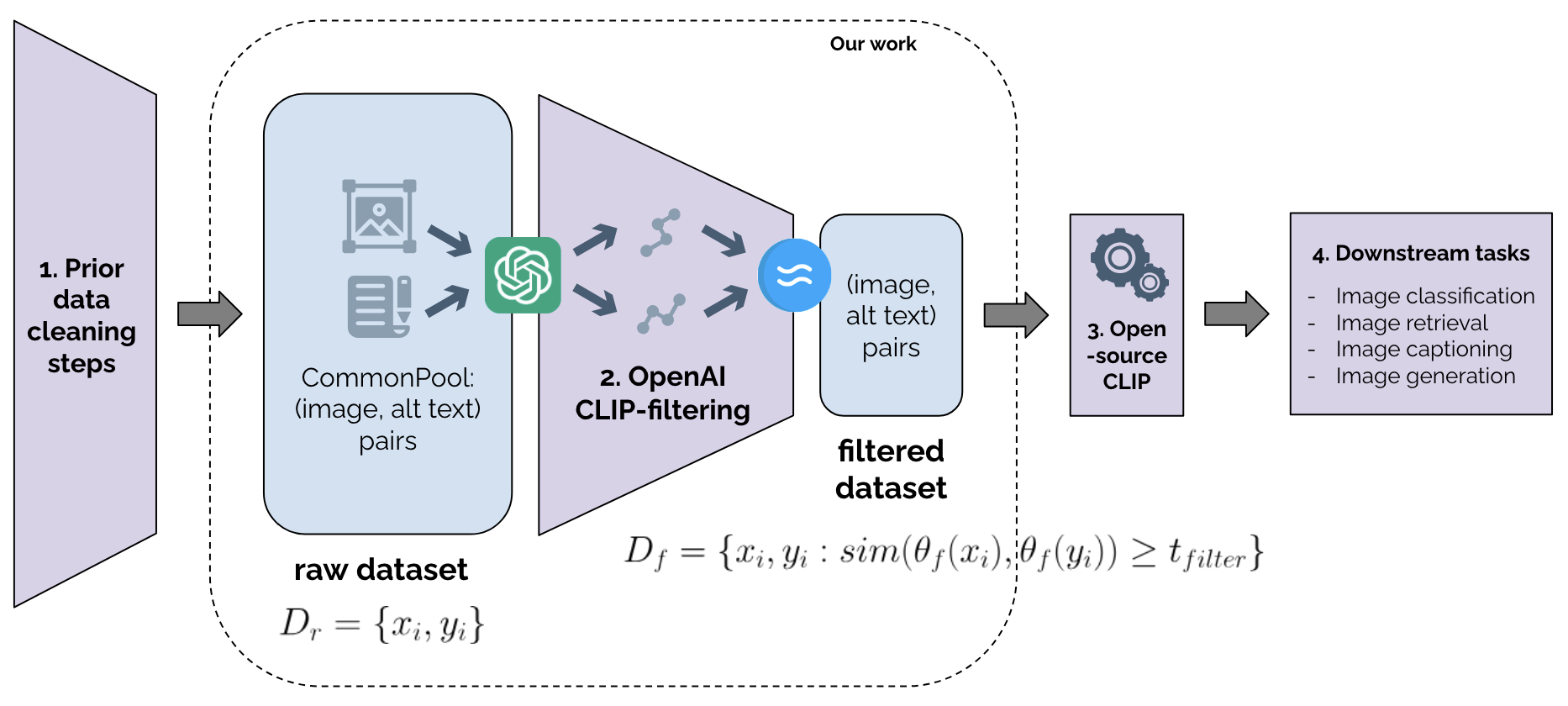}
    \end{center}
    \caption{CLIP-filtering pipeline of LAION \citep{laion, laion5b} and DataComp \citep{datacomp} as described in \Cref{sec:bkgd:pipeline}. In step 1, a series of initial data cleaning techniques is applied to CommonCrawl \citep{commoncrawl} to form the raw dataset $D_r$ with each sample as image $x_i$ and corresponding alt-text tag $y_i$. The filtered dataset $D_f$ is obtained by step 2, which applies the pre-trained OpenAI CLIP model $\theta_f$ to each image-text pair $(x_i, y_i)$. If the cosine similarity score between embeddings $\theta_f(x_i)$ and $\theta_f(y_i)$ is above some predefined threshold $t_{filter}$, then the pair is included in the filtered dataset. From this dataset, step 3 trains the open-source CLIP model, and step 4 applies the CLIP model to various downstream tasks. We scope our investigation to step 2 within the enclosed box.}
    \label{fig:pipeline}
    \Description{Pipeline of LAION dataset in stages: Step 1 is a funnel that represents prior data cleaning steps applied to obtain the raw dataset, which is composed of image, alt-text pairs. Step 2 is another funnel that applies OpenAI CLIP-filtering based on the CLIP embeddings of image and text, in order to create the filtered dataset. Step 3 trains the open-source CLIP model, and Step 4 trains the downstream tasks, which include image classification, image retrieval, image captioning, and image generation. There is a rectangular box that encloses the raw dataset, Step 2, and the filtered dataset, which refers to the scope of our work.}
\end{figure*}

To address these issues, machine learning practitioners and researchers have increasingly begun to rely on automated data filtering to improve training efficiency and discard what they refer to as ``low-quality'' data. Specifically, for multimodal data, researchers have applied a pretrained CLIP model to filter image-text data obtained from Common Crawl. As depicted in \Cref{fig:pipeline}, this \textit{CLIP-filtering} method (referring to the use of the OpenAI CLIP model \citep{clip}) assesses similarity between an image and its corresponding text within the model's embedding space. This filtering technique was used to create the LAION-400M \citep{laion}, LAION-5B \citep{laion5b}, and DataComp-1B \citep{datacomp} datasets with sizes of 400 million, 5 billion, and 1 billion respectively, which were then used to train other open-source CLIP models. Because these models obtained state-of-the-art performance on several zero-shot image classification tasks, researchers have concluded CLIP-filtering to be the ``most performant'' method, as opposed to other filtering techniques that use caption length or image size \citep{datacomp, fang_data_nodate}.

However, OpenAI CLIP is neither intended for filtering nor built for deployment \citep{birhane}. As stated in their model card \citep{modelcard}: ``Any deployed use case of the model --- whether commercial or not --- is currently out of scope.''
Here, using CLIP as a filter to build a deployed model is by definition a \textit{use} case of CLIP in a deployed setting.
Yet these datasets obtained via CLIP-filtering have subsequently been used to train text-to-image models like Stable Diffusion and Midjourney, which each has tens of millions of users \citep{stablediffusion, midjourney}. Recent work has shown that these models exhibit problematic behaviors, such as amplifying demographic stereotypes or generating violent and sexually-explicit content \citep{bianchi_easily_2023}. Much of these behaviors is attributed to the content of the training data \citep{unsafe}, as prior work demonstrates that the LAION datasets contain high rates of hateful content and misogynistic stereotypes \citep{den}.

At the same time, it is unclear what role CLIP-filtering plays on downstream text-to-image models or whether CLIP-filtering simply replicates the model's training dataset, especially given the demographic biases embedded in OpenAI CLIP itself \citep{agarwal_evaluating_2021}. While there have been speculations on the demographic biases of CLIP-filtering \citep{birhane}, there has not been any exploration to assess exactly how the filter changes the demographic makeup of the training dataset. As such, we choose to examine this initial stage of dataset curation because of its potential impact on downstream models.

\subsection{Contributions}
\label{intro:contributions}

In our work, we perform an audit of the standard CLIP-filtering approach to the DataComp CommonPool dataset, focusing specifically on the use of the OpenAI CLIP model as a filter, as a case study of the role data filtering plays within the broader machine learning pipeline. To guide our analysis, we seek to answer the following question: \textbf{In what ways does CLIP-filtering exclude or include various types of data in DataComp CommonPool?} We focus on this dataset specifically because of its usage as an academic benchmark, although our findings may extend to LAION-400M and LAION-5B since they were created in a similar manner.

We break down our investigation to address the following three questions: First: \textbf{Which demographic groups are disproportionately excluded by CLIP-filtering?} Different demographic groups being filtered at different rates implies a form of representational harm \citep{shelby2023sociotechnical}. A filter that systematically omits data from a certain group may therefore worsen the performance of the resulting model on that group. As a result, we aim to quantify this breakdown between unfiltered and filtered data across various demographic dimensions. The need to assess sociodemographic attributes across multiple modalities motivates our next question: \textbf{How can we measure sociodemographic attributes at scale within an image-text dataset?} We examine a broad range of factors like gender, age, race, religion, sexuality, language, or geographic region. While by no means comprehensive, we infer attributes that appear in the text and image components that can easily be related to people. In addition to examining filtering discrepancies by demographics, we also examine discrepancies by data source: \textbf{What types of websites are considered ``high-quality'' by the CLIP filter and what are the implications of certain websites being included?}
To the extent that it is valuable to know whether datasets are globally representative, we ask whether data from certain geographic regions or time periods of the internet are treated differently by CLIP-filtering, as well as what types of websites have data that are more likely to pass through the filter.

Driven by the above questions, we conduct a novel examination of demographic bias as a consequence of multimodal filtering. In doing so, we make the following contributions:

\begin{enumerate}
    \item \textit{Filter discrepancies:} Across multiple modalities, different imputed sociodemographic groups are filtered at different rates. For example, CLIP-filtering is more likely to exclude data relating to LGBTQ+ identities and non-Western regions.
    \item \textit{Exclusion amplificaton:} We find that not only does CLIP-filtering disproportionately exclude data from certain imputed demographic groups, but that this form of exclusion \textit{amplifies} existing representation disparities. Some imputed groups that are already underrepresented in the original dataset are disproportionately filtered out at higher rates compared to overrepresented groups.
    \item \textit{Website inclusion:} Data from stock photo websites and U.S.\ and British news sites, which may be subject to copyright restrictions, are included at much higher rates than average. Large quantities of images from websites serving sexually-explicit material are also present after both NSFW and CLIP-filtering.
    \item \textit{Evaluation tools:} To address our research questions, we also develop novel methodologies to effectively audit an image-text dataset at scale. We make our analysis code available at \url{https://github.com/hongrachel/clip-filtering-bias/}.
\end{enumerate}

Overall, this existing data filtering method is a clear example of a value-embedded machine learning practice \citep{values}. From our results, we see that CLIP-filtering does in fact substantially change the makeup of the final dataset. This reinforces more broadly the idea from \citet{suchin} that data filtering must encode a specific ideology of what constitutes as ``high-quality data.'' Prior works on image-text filtering assume that ``quality'' is often intrinsic to the data and that this ``noise'' to be weeded out is vaguely defined \citep{wenzek2019ccnet}. The LAION and DataComp authors demonstrate the efficacy of the CLIP filter by selecting specific downstream metrics for specific datasets such as zero-shot ImageNet classification accuracy, rather than the societal impacts of downstream models \citep{laion, datacomp}. Here, we investigate a case where filtering is designed around particular assumptions of what data quality and performance should be without additional justification. \textit{Despite the supposed objectivity of these decisions, our findings prove that filtering embeds societal norms as to what should and should not be excluded}. Following our analysis, we further elaborate on these implications and give recommendations for designing data filters in the future.

\section{CLIP-filtering}
\label{sec:bkgd}

We describe the filtering process first proposed as a high-performing filtering method to curate the LAION datasets \citep{laion, laion5b} and subsequently studied and expanded upon in DataComp \citep{datacomp}. 

Broadly, the goal of the image-text filtering step is to determine whether a piece of text accurately describes its corresponding image --- in other words, to obtain image-text alignment. However, the term ``alignment'' is not well-defined nor well-measured, and it is unclear how using CLIP fits with the search for ``high-quality'' data.

\subsection{Machine learning pipeline}
\label{sec:bkgd:pipeline}

\Cref{fig:pipeline} provides an overview of the machine learning pipeline to build the LAION and DataComp datasets, including the CLIP-filtering step, in order to train large image-text models on these datasets. In our work, we primarily follow the DataComp pipeline as each step is well-documented, although this process is easily extendible to the LAION pipeline.

\subsubsection{Raw dataset collection}

Image-text data is first scraped from a snapshot of the web, CommonCrawl \citep{commoncrawl}, through parsing HTML image tags with nonempty alt-text tags. Alt-text, or alternative text, is a description associated with an image when the image cannot be rendered. Primarily, alternative text is intended for accessibility purposes, such as being presented to users with screen readers \citep{alttext}. However, extensive research demonstrates that a large proportion of websites do not follow standard accessibility guidelines for alt-text \citep{webinsight}. For example, among the one million most popular websites, $22.1\%$ of home page images lack alt-text tags, and $10.9\%$ of images with alternative text were non-descriptive including words like ``text'' or ``blank'' \citep{webaim}. In addition, alt-text is a primary way search engines parse images, leading many alt-text tags being designed to improve website visibility on search engines rather than for accessibility or fidelity \citep{guinness2018caption}. As a result, in order to build high-quality image-text datasets from the web, it becomes necessary to filter instances with questionable alt-text.

Once the dataset curators gather the image URLs and alt-text pairs, they attempt to download the images and keep the image-text pairs that are successfully downloaded. Some data cleaning methods are also applied here in the DataComp pipeline: they remove exact url-text duplicates within the dataset as well as potential image overlap with pre-selected evaluation sets. In addition, LAION removes short text and extremely small or large image sizes, while DataComp applies several toxicity classifiers to discard NSFW-detected images or text. This toxicity filtering is another key component of data filtering, and prior work has demonstrated that current hate speech detection models are more likely to label text written by African Americans as offensive \citep{toxic-bias}. However, we limit the scope of our work to the CLIP-filtering step (i.e., after the NSFW filter has been applied), although the toxicity filtering step may also disproportionately exclude data from certain identities.

After these preprocessing steps, the authors of DataComp name the subsequent dataset as \textbf{CommonPool} which consists of 12.8 billion pairs. For LAION-5B, this results in a dataset of 50 billion pairs before CLIP-filtering \citep{laion5b}. For the sake of clarity, we refer to these image-text pairs before CLIP-filtering (CommonPool in the case of DataComp) as the \textbf{raw dataset}.

\subsubsection{CLIP-filtering}

After the creation of the raw dataset, the main filtering step uses a pre-trained CLIP model (ViT-B/32 or ViT-L/14) as follows: for every image-text pair, extract the CLIP image and text embeddings and obtain the cosine similarity score between these embeddings. The image-text pair passes the filter if the score is above a predefined threshold and discarded otherwise.

Each prior work uses differently-chosen similarity score thresholds --- LAION-400M \citep{laion} has a threshold of 0.3, LAION-2B-en (the subset of LAION-5B with English-detected text) \citep{laion5b} has a threshold of 0.28, and DataComp has the threshold that separates the top 30\% of data in CommonPool (0.281 for ViT-B/32 and 0.243 for ViT-L/14) \citep{datacomp}. Across all of these works, this discards a majority of the data from the raw dataset, and this threshold is chosen in order to optimize a specific set of downstream metrics.
We refer to the downstream dataset obtained by applying the CLIP filter as the \textbf{filtered dataset}.

We note that for DataComp, the filtering task on the CommonPool benchmark is a track in their proposed competition in order encourage better filtering models \citep{datacomp_track}. CLIP-filtering is only one of the many filtering baseline experiments they conduct, but they find that their best baseline to improve their evaluation tasks, including ImageNet zero-shot accuracy, incorporates CLIP-filtering \citep{datacomp}. Thus, the CLIP-filtering step is a key component of the DataComp-1B dataset they release.

\subsubsection{Pre-training}

The filtered dataset then becomes the training dataset for any large model across a variety of tasks. The LAION and DataComp researchers focus on training open-source CLIP models following the original CLIP paper \citep{openclip, clip}. At the same time, these filtered datasets have also been used to train large text-to-image models like Stable Diffusion or Midjourney, despite OpenAI and LAION stating they are not ready to be used for commercial purposes \citep{modelcard, stablediffusion, midjourney}. Given that \citet{datacomp} demonstrate that various filtering methods impact the resulting CLIP model's performance, it becomes evident that filtering plays an important role in developing pre-trained models, whether for research or for deployed purposes. As a result, this CLIP-filtering approach has curated datasets to train models with a massive reach over tens of millions of users \citep{stablediffusion, midjourney}.

\subsubsection{Fine-tuning on downstream tasks}

The final step before model deployment can include fine-tuning or training another model, such as a GAN \citep{clip-imagecaptioning}, using the pre-trained CLIP model from the previous step. In the next section, we describe how CLIP can be used for various downstream tasks, such as image classification or image retrieval.

\subsection{CLIP model}
\label{sec:bkgd:clip}

Contrastive Language-Image Pretraining, or \textit{CLIP}, was proposed by \citet{clip} as a pre-trained model that learns images from natural language text through mapping images and texts to the same representation space. The training objective follows contrastive representation learning \citep{contrastive}: to maximize the cosine similarity between image embeddings and their corresponding text embeddings, while minimizing similarity between all other incorrect pairings. The original CLIP model was trained on WebImageText, which is an unreleased dataset of 400 million English-text pairs curated with 500,000 web queries. Subsequent work has followed similar model architecture to train open-source versions of CLIP on publicly-available datasets. \citet{metaclip} replicate the training data for the original CLIP model by reconstructing the query metadata, whereas \citet{openclip} train on the LAION dataset, which imitates WebImageText but still relies on the original model as a filter.

The authors of the original CLIP model \citep{clip} and following work demonstrate that CLIP obtains state-of-the-art performance on various vision tasks without additional training, such as zero-shot ImageNet classification, image retrieval \citep{ecommerce-retrieval, sketch-retrieval}, as well as using the embeddings to guide image captioning \citep{clip-imagecaptioning} and text-to-image models \citep{clip-generation}.


\section{Related Work}
\label{sec:related_work}

In this section, we highlight previous work on bias audits and data filtering methods that are relevant to our study. We aim to contribute to the existing discussion on data filtering bias through a novel case study of data filtering on a popular large-scale dataset. 

\subsection{CLIP bias}
\label{sec:related_work:clip_bias}

Since the release of the OpenAI CLIP model, multiple follow-up investigations have found that CLIP encodes harmful stereotypes and sociodemographic biases which may have been present in the training data. Notably, \citet{agarwal_evaluating_2021} find that CLIP misclassifies images of Black people as non-human at higher rates compared to images of other racial groups, as well as images of men as `executive' or `doctor' at higher rates compared to images of women. \citet{wolfe_american_2022} also find that CLIP embeddings of images of white people are more associated with being ``American.'' Other works evaluate CLIP-aided image retrieval and text-to-image models and demonstrate representation bias as well as amplification of stereotypes \citep{ali_evaluating_2023, bianchi_easily_2023}. As a result, we hypothesize that CLIP-filtering may further overrepresent content containing stereotypes or harmful biases and disproportionately exclude content that do not follow these problematic assocations.

\subsection{CLIP-filtered dataset audits}
\label{sec:related_work:audit}

Some recent works have audited the LAION datasets in order to understand the nature of the content within these large-scale datasets. Recently, \citet{thiel_identifying_nodate} uses hash-based detection and other computational techniques to identify thousands of CSAM (Child Sexual Abuse Material) images in LAION-5B, which resulted in the takedown of the LAION-5B dataset in December, 2023 \citep{csam}. In addition, \citet{birhane, den} have found significant rates of hateful content, explicit images, stereotypes, and racial slurs in both the LAION-400M and LAION-2B-en datasets. Given the biases of CLIP, they illustrate hypothetical examples in which misogynist or racist descriptions would pass the CLIP filter but reasonable, benign descriptions would be excluded \citep{birhane}. The question then remains: What, exactly, does CLIP-filtering include and exclude? We extend their findings through examining the impact of filtering on demographic representation at scale.

\subsection{Data filtering bias}
\label{sec:related_work:filter_bias}

In NLP, filtering is a common practice to build large text datasets scraped from the internet. For instance, to curate the training dataset for the GPT3 model, researchers trained a quality filter to identify high-quality sources like Wikipedia \citep{gpt3}. \citet{suchin} find that the GPT3 quality filter is more likely to classify text from wealthier, urban, and larger schools as high-quality. \citet{c4} also examine the blocklist filter used to create the popular C4 dataset \citep{c4t5} and determine that mentions of identities from marginalized groups are more likely to be filtered out. \citet{aboutme} investigate many English language and quality text filters applied to desecriptions of website creators and show disparate rates of filtering based on topic and geographic region. For tabular data, common data preprocessing steps are found to remove data from historically disadvantaged groups and worsen downstream model fairness \citep{biswas_fair_2021, guha_automated_2023}. 

There also has been much discussion on the systematic exclusion of data from marginalized identities in the data cleaning stage more broadly. \citet{parrots} caution early on that filtering can suppress discourse from marginalized identities and compound power imbalances in text data collection practices. From a sociotechnical lens, \citet{muller2022forgetting} describe how data cleaning can be a method of erasure that is difficult to detect. \citet{archival} draw on an archival perspective to examine the social value judgements embedded in filtering choices. We are inspired by prior filtering discussions and extend their approaches to a commonly used multimodal data pipeline. To our knowledge, there has been no work assessing data filtering bias at scale in the context of image-text datasets.

\subsection{Multimodal data filtering}
\label{sec:related_work:multimodal}

Researchers have also investigated image-text data filtering in relation to performance or robustness to distribution shifts. \citet{nguyen2022quality} demonstrate that filtering noisy data with a pre-trained robust model like CLIP can lead to more robust models downstream, which may explain the state-of-the-art image classification accuracy by training on the LAION and DataComp datasets. \citet{fang_data_nodate} complicate this result by showing that using a higher-performing CLIP model as a filter counterintuitively does not always lead to better downstream models. While these findings reveal the intricacies of building effective multimodal filters, the ongoing work so far has not yet focused on societal implications.

\subsection{Downstream text-to-image model harms}
\label{sec:related_work:downstream}

There has been ample work that evaluate the societal issues of downstream models trained on CLIP-filtered data. \citet{luccioni2024stable}, for instance, find that Stable Diffusion (which trains on LAION-5B obtained via CLIP-filtering \citep{Rombach_2022_CVPR}), underrepresents marginalized identities in their image generations. Moreover, researchers have demonstrated that Stable Diffusion generates unsafe, hateful, or sexualized content at significantly high rates \citep{unsafe, wolfe2023contrastive}. Exposure to stereotypical and problematic imagery in general has been shown to shape people's beliefs and behavior \citep{branley2017exposure, slusher1987reality}, and this becomes more pressing as these models grow increasingly popular. As such, it thus becomes necessary to understand exactly where societal harms may arise in the ML pipeline.

\section{Approach}
\label{approach}

We first replicate the CLIP-filtering step for the \texttt{small} version of CommonPool of 12.8 million image-text pairs. We use the CLIP ViT-L/14 model and set the threshold as $0.243$ to select the top $30\%$ of pairs according to CLIP similarity score as done in prior work \citep{laion5b, datacomp}, resulting in a filtered dataset with 3.84 million samples. In this section, we describe how we obtain demographic annotations and measure filtering discrepancies.

\subsection{Imputed demographic group annotations}
\label{approach:demographic_group}

Many large datasets, including multimodal datasets like CommonPool, contain little or no metadata about the people used in generating their data, which can include people in images, the authors of text, or patients whose medical records comprise a dataset. This makes auditing datasets and their downstream uses for disparate treatment along certain characteristics, like gender, race, and age, a challenging task. Because CommonPool does not contain demographic metadata, we cannot directly audit CLIP-filtering along these axes. Instead, we attempt to audit CLIP-filtering's interaction with \textit{imputed} demographic information, using existing models to evaluate filtering bias.

We determine demographic group or geographic region across multiple modalities in the CommonPool dataset: from the text content, the image content, and the source URL of the sample. As described in \Cref{sec:demographic_group}, this includes methods such as searching for mentions of keywords relating to demographic identity or applying an external face attribution predictor. Due to the sample size considered, we separate our analysis by modality, despite a sample containing both image and text, and leave the image-text intersection for follow-up work. We emphasize these findings \textit{do not represent behavior of filtering according to true demographics}; instead, they reveal how filtering interacts with imputed demographic attributes. While we elaborate on the limitations of imputed demographics in \Cref{sec:approach:limits}, findings here still have broad implications, as imputed demographics can correlate, although imperfectly, with socially salient populations \citep{elliott2009using}.

\subsection{Evaluation metrics}
\label{sec:approach:eval}

To assess the impact of filtering, we track the \textbf{pass rates} by imputed demographic group, which refer to the percentage of raw dataset that passes through the filter. In other words, a \textit{higher pass rate} means more data is kept in the filtered dataset, and a \textit{lower pass rate} means more data is excluded from the filtered dataset. This allows us to evaluate whether CLIP-filtering leads to representational harms in the filtered dataset if data relating to certain imputed demographic groups are considered ``lower quality'' on average.

\subsection{Limitations}
\label{sec:approach:limits}

We recognize numerous limitations to the sociodemographic imputation and source analysis techniques that we use. Third-party gender predictors, for instance, have been demonstrated to have disparate error rates across sensitive attributes and can rely on biases present in the face detection step \citep{gendershades, rekogbias}. Furthermore, these predictors only give binary labels as ``Male'' and ``Female'' which conflates gender with sex and ignores non-binary individuals \citep{valdes1996unpacking}. Identity keywords in the text also are rough proxies for the presence of sociodemographic attributes in the image or text. We acknowledge that gender and race are not inherently visual components, but rather socioculturally defined \citep{gender, race}. 

Evaluating filtering according to these imputed demographic groups is a form of measuring filtering's group fairness properties. Group fairness is not sufficient especially when individuals belong to multiple groups or their group categories are not explicitly known \citep{dwork}. At the same time, we recognize the difficulties of assessing sociodemographic information at scale --- we believe that determining relationships between pass rates and these signals, although problematic or noisy, reveals behavior of the CLIP filter and the content of data in the LAION and DataComp-1B datasets.

Moreover, while any difference in pass rates that we find can be attributed to the impact of CLIP-filtering, it does not necessarily demonstrate a causal relationship that a certain demographic group will always be excluded by CLIP. There may be underlying differences in the raw dataset, where data relating to a particular group may come from websites that have established alt-text practices, yet this is infeasible to assess at scale across CommonPool. We argue regardless that disparities in filtering still matter in order to expose whether certain imputed demographic groups are being omitted in the construction of large-scale machine learning datasets. Thus, it still remains important to understand the impact filtering has on the resulting dataset given that these CLIP-filtered datasets like LAION and DataComp are used widely in both industry and research \citep{stablediffusion, midjourney, datacomp}.

\section{Demographic Group Analysis}
\label{sec:demographic_group}

This section describes the demographic group imputation methods along both text and image modalities and presents their corresponding results. We include additional analysis in \Cref{appx:results}.

\subsection{Identity keyword}
\label{sec:demographic_group:identity}

\paragraph{Method: } We first identify mentions of demographic groups through simple keyword search, by following the same methodology as \citet{c4}, which focuses on blocklist filtering of text. Their list of regular expressions (shown in \Cref{tab:keywords}) investigates demographic dimensions related to gender, sexual orientation, race, and religion --- while no means comprehensive, this initial exploration of filtering analysis allows us to compare our results to their findings on the blocklist filter for the C4 text dataset \citep{c4t5}. Next, we plot the pass rates (rate of inclusion in the filtered dataset) grouped by mentions of identity keywords in the text modality. Upon manual inspection, we dismiss analysis of the \texttt{white} and \texttt{black} keywords since text with these keywords typically describe clothing apparel rather than people.

\begin{table}[t]
\caption{List of regular expressions relating to identity keywords used in \citet{c4}.}
\label{tab:keywords}
\begin{center}
\begin{tabular}{lll}
\hline
\texttt{african[ -]americans?} &
\texttt{asian([ -]american)?s?} \\
\texttt{bi-?sexuals?} &
\texttt{blacks?} \\
\texttt{caucasians?} &
\texttt{christians?} \\
\texttt{european([ -]american)?s?} &
\texttt{females?} \\
\texttt{gays?} &
\texttt{heterosexuals?} \\
\texttt{homosexuals?} &
\texttt{jew(|s|ish)?} \\
\texttt{latin[oax]s?} &
\texttt{lesbians?} \\
\texttt{m[ae]n} &
\texttt{males?} \\
\texttt{muslims?} &
\texttt{non[-]?binary} \\
\texttt{straights?} &
\texttt{trans(|\textbackslash+|gender)} \\
\texttt{whites?} &
\texttt{wom[ae]n}
\\ \hline
\end{tabular}
\end{center}
\end{table}

\subsubsection{The CLIP filter excludes data relating to LGBTQ+ identities at higher rates.}
\label{sec:demographic_group:identity:lgbtq}

\begin{figure}
    \centering
    \begin{subfigure}[b]{\linewidth}
         \includegraphics[width=\linewidth]{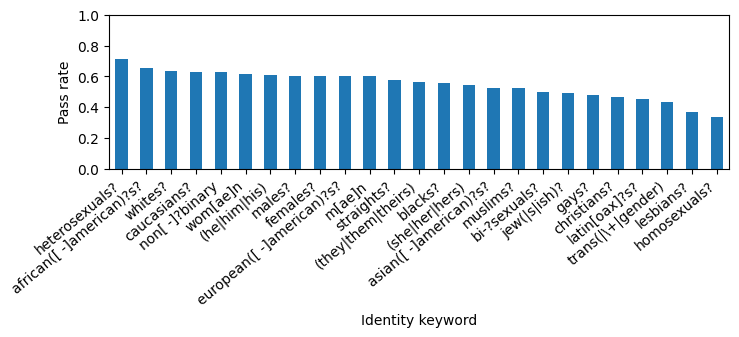}
        \caption{Pass rate by identity keyword sorted in descending order.}
        \label{fig:identity_keyword_ratios}
    \end{subfigure}

    \begin{subfigure}[b]{\linewidth}
        \includegraphics[width=\linewidth]{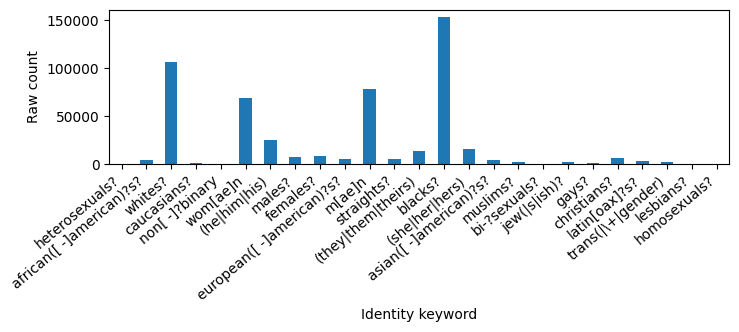}  
        \caption{Frequency in raw dataset by identity keyword.}
        \label{fig:identity_keyword_raw_total}
    \end{subfigure}
    
    \caption{Pass rate (a) and raw dataset frequency (b) broken down by mentions of identity keywords from \Cref{tab:keywords}. A higher pass rate represents a higher proportion included in the resulting CLIP-filtered dataset. While \texttt{white} and \texttt{black} terms are commonly mentioned, we inspect samples manually and find they relate overwhelmingly to clothing items. Figure (a) shows that CLIP is more likely to exclude text samples containing LGBTQ+ keywords compared to other identity keywords.}
    \label{fig:identity_keyword}
    \Description{The top subfigure is a bar graph that plots pass rates for each identity keyword from \Cref{tab:keywords}. Pass rates vary from $0.7$ to $0.3$, where \texttt{gay}, \texttt{Christian}, \texttt{Latinx}, \texttt{transgender}, \texttt{lesbian}, and \texttt{homosexual} keywords have the lowest pass rates. The bottom subfigure is a bar graph that plots the raw count by identity keyword in the same order. The plot shows varying raw counts by identity keyword, but very low count for \texttt{heterosexual}, \texttt{nonbinary}, \texttt{bisexual}, \texttt{lesbian}, \texttt{homosexual} keywords.}
\end{figure}

In \Cref{fig:identity_keyword}, we see similar trends as \citet{c4} found on text filtering, where mentions of keywords relating to LGBTQ+ identities (\texttt{homosexual}, \texttt{lesbian}, \texttt{transgender}, \texttt{gay}, \texttt{bisexual}) are excluded at much higher rates compared to mentions of other keywords. We extend existing findings on text to the multimodal setting: Cleaning methods often remove data relating to LGBTQ+ identities, although prior work specifically examines how hate speech classifiers are more likely to label this type of data as toxic \citep{lgbtqtoxic}. This difference is notable given that CLIP is not trained explicitly as a toxicity classifier, yet as a filtering method still obtains the same trends.

\subsubsection{Intersections in identity keywords reveal additional filtering discrepancies.}
\label{sec:demographic_group:identity:intersections}

\begin{figure}
    \centering
    \includegraphics[width=\linewidth]{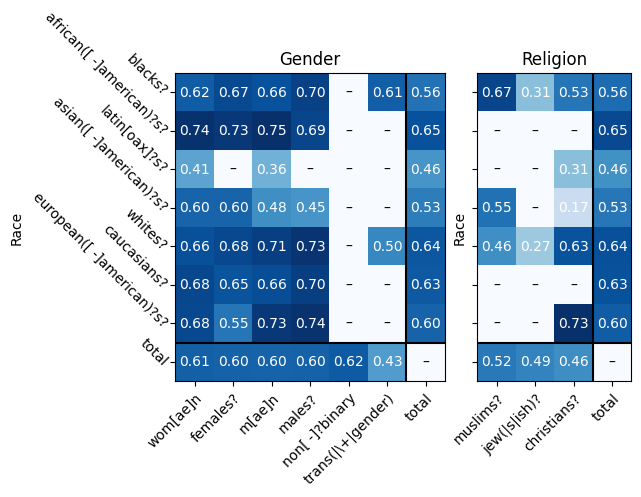}
    \caption{Heat map of pass rates by intersections of various demographic dimensions where a higher pass rate is a darker shade of blue. ``---'' indicates that the raw frequency of samples within that intersection is below 10 and therefore not reported. The ``total'' row and column in each map refer to the pass rate for all samples that mention one identity keyword. For instance, the \texttt{total}, \texttt{wom[ae]n} square shows that 61\% of samples that mention keyword \texttt{wom[ae]n} pass the CLIP filter.}
    \label{fig:intersection}
    \Description{Heat map of pass rates for intersections of gender and race-related keywords on the left and religion and race-related keywords on the right. Due to low sample size of the intersecting words, \texttt{nonbinary}, \texttt{transgender}, \texttt{Muslim}, and \texttt{Jewish} keywords are mostly omitted when intersecting with race-related keywords. \texttt{Latinx} and \texttt{Asian}-related keywords have pass rates of $0.41$ and $0.60$ respectively when intersected with woman-identifying terms compared to pass rates of $0.36$ and $0.48$ respectively for man-identifying terms. The \texttt{European} keyword has a pass rate of $0.73$ when intersected with man-identifying terms compared to a pass rate of $0.68$ for woman-identifying terms. For race and religion intersections, we find a pass rate of $0.17$ for the \texttt{Asian} and \texttt{Christian} intersection, and a pass rate of $0.73$ for the \texttt{European} and \texttt{Christian} intersection.}
\end{figure}

We follow up this analysis and examine intersections between various dimensions of sociodemographic identity. This limits our sample size, so in \Cref{fig:intersection} we only include intersections with frequency in the raw dataset of at least 10. We observe that \texttt{Latinx} and \texttt{Asian}-related keywords have higher pass rates when intersected with woman-identifying terms compared to man-identifying terms, while the \texttt{European} keyword has higher pass rates when intersected with man-identifying terms compared to woman-identifying terms. For race and religion intersections, we find relatively low pass rates for the \texttt{Asian} and \texttt{Christian} intersection, and relatively high pass rates for the \texttt{European} and \texttt{Christian} intersection, compared to the average pass rates for a single keyword in the row or column of \Cref{fig:intersection}.

\subsubsection{Common words associated with certain genders embed gender stereotypes.}
\label{sec:demographic_group:identity:word_stereotypes}

\begin{figure}
    \begin{center}
    \includegraphics[width=\linewidth]{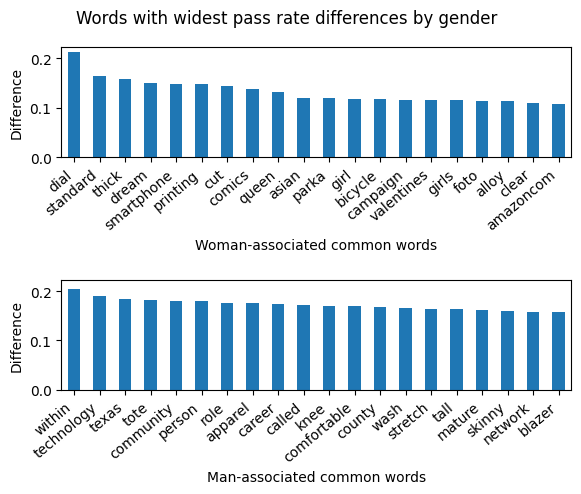}
    \end{center}
    \caption{Common words with widest pass rate differences by gender. The top graph plots the top 20 words with the largest pass rate difference between mentions of women keywords versus mentions of men keywords and hence are more ``woman-associated.'' The bottom graph plots the top 20 words with the largest pass rate difference between men and women and hence are more ``man-associated.''}
    \label{fig:common_gender}
    \Description{Two bar graphs of common words with widest pass rate differences between women and men. The top bar graph plots the pass rate disparity for the top 20 common words with higher pass rates for women, and the bottom bar graph the pass rate disparity for the top 20 common words with higher pass rates for men. For both graphs, the difference in pass rate ranges from $0.1$ to $0.2$. The common words with significantly higher pass rate for women than men include words like \texttt{queen}, \texttt{asian}, \texttt{girl}, and \texttt{valentines}. The common words with significantly higher pass rate for women than men include words like \texttt{technology}, \texttt{texas}, \texttt{person}, \texttt{career}, \texttt{comfortable}, \texttt{tall}, and \texttt{mature}. }
\end{figure}

To examine associations between words and mentions of gender, we isolate text samples that mention man or woman-related keywords (listed in \Cref{sec:appx:methods:gender_keywords}). Of those samples, we find a list of common words that appear in at least $100$ samples, ignoring a set of stop words from the \texttt{word\_cloud} library \citep{word_cloud}.
For each common word $w$, we calculate the pass rate for samples that contain both $w$ and a woman-related keyword, and for samples that contain both $w$ and a man-related keyword. In \Cref{fig:common_gender}, we plot the common words with the largest pass rate difference between woman and man-related samples --- the top figure shows words with substantially higher woman-related pass rates, and the bottom figure shows words with substantially higher man-related pass rates.
Common words that have a relatively higher woman-related pass rate are stereotypically associated with women: for instance, \texttt{queen}, \texttt{asian}, \texttt{girl}, and \texttt{valentines}. Common words that have a relatively higher man-related pass rate include \texttt{technology}, \texttt{texas}, \texttt{person}, \texttt{career}, \texttt{comfortable}, \texttt{tall}, and \texttt{mature}.

The aforementioned gender stereotype associations provide evidence that CLIP-filtering ranks data according to a stereotypical form of \textit{alignment} between the text and images, rather than some inherent measure of their \textit{quality}. In this example, a common word paired with one gender is included by the filter more often than the same word paired with a different gender. We see the feminization of the Asian identity \citep{asian} and the diminution of women as ``girls'' \citep{girl}. We also observe that text containing the word ``person'' is more likely to pass the filter when paired with words related to men than when paired with words related to women. Other common words with higher man-related pass rates illustrate the ongoing stereotypes of associating men with technology and careers, which have been shown to be present among both humans and machine learning models~\citep{oldenziel1999making, zuo2000breadwinner, caliskan2022gender}.

\subsection{Image group}
\label{sec:demographic_group:image_group}

In this section, we describe results from analysis of the image component of the CommonPool dataset. Overall, we show that images of various imputed demographic groups according to gender, age, and race are filtered at different rates. Because image demographic imputation methods can be inaccurate for particular demographic groups \citep{gendershades}, we apply two imputation techniques: (1) an external face attribute predictor Amazon Rekognition \citep{rekog} and (2) a novel kNN-based clustering method of CLIP embeddings.

\paragraph{Gender and age annotation method:} We pass a subsample of $100,000$ CommonPool images through Amazon Rekognition \citet{rekog} in order to obtain face detections as well as imputed gender and age annotations on those detected faces. Out of this subsample, Rekognition detects faces contained in about $18,000$ images, and to disambiguate between different people, who might have distinct imputed demographic attributes, we perform analysis on the $11,000$ of those images that contain only one detected face. In this section, for consistency we refer to Rekognition gender labels of \texttt{Male} and \texttt{Female} but note that they conflate gender with sex \citep{valdes1996unpacking}. We recognize the numerous issues and biases surrounding face detection and face attribute prediction as mentioned in \Cref{sec:approach:limits}.

\subsubsection{Rekognition classifies more images as \texttt{Male} than \texttt{Female} in the raw dataset, and this gap widens after the CLIP filter}

Out of the $11,000$ images that are detected by Rekognition as containing a single face, there are more images with a \texttt{Male}-imputed face than images with a \texttt{Female}-imputed face, and the pass rate for the \texttt{Male}-imputed group is about $3.8$ percentage points higher than the pass rate for the \texttt{Female}-imputed group. This difference in pass rate thus widens the representation gap after filtering --- for Rekognition-annotated images, there initially are 42.1\% more images in the \texttt{Male}-imputed group than images in the \texttt{Female}-imputed group in the raw dataset, but after filtering this disparity jumps to 63.9\%. 

This result demonstrates a notion of bias we call \textit{exclusion amplification}. Data from imputed groups already underrepresented in the raw dataset are excluded at even higher rates (i.e. have a lower pass rate) compared to data from overrepresented groups --- in other words, representation bias is amplified after filtering.

\subsubsection{When split by imputed gender and age annotations, highly represented groups also pass through the filter at higher rates compared to less represented groups.}

\begin{figure}
    \begin{center}
    \includegraphics[width=\linewidth]{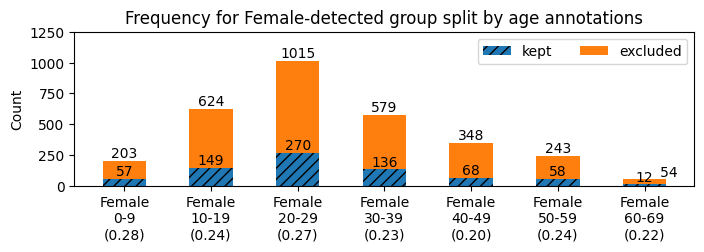}
    
    \includegraphics[width=\linewidth]{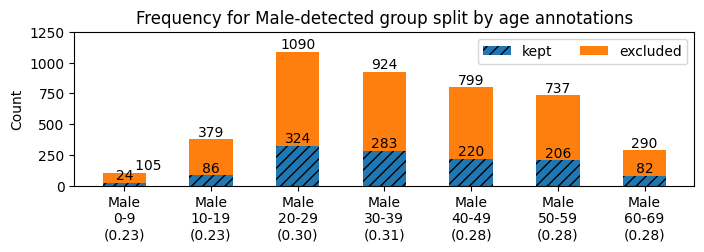}
    \end{center}
    \caption{Frequency for samples detected by Rekognition as containing one face, split by Rekognition-detected age and gender groups. For each imputed group the pass rate is included in parentheses. We see that older detected ages in the \texttt{Female}-imputed group have substantially lower pass rates and lower representation than the corresponding \texttt{Male}-imputed group, and that younger detected ages in the \texttt{Male}-imputed group have lower pass rates and lower representation than the corresponding \texttt{Female}-imputed group.}
    \label{fig:age_gender}
    \Description{Bar graphs of frequency of kept and included samples split by Rekognition-detected age and gender groups. The top bar graph shows frequency for the \texttt{Female}-imputed group, and the bottom bar graph shows frequency for the \texttt{Male}-imputed group. Older detected ages between $40-69$ in the \texttt{Female}-imputed groups have pass rates of $0.20, 0.24, 0.22$ and overall frequency of $348, 243, 54$, while the corresponding \texttt{Male}-imputed groups have pass rates of $0.28, 0.28, 0.28$ and overall frequency of $799, 737, 290$. Younger detected ages between $0-19$ in the \texttt{Female}-imputed groups have pass rates of $0.28, 0.24$ and overall frequency of $203, 624$, while the corresponding \texttt{Male}-imputed groups have pass rates of $0.23, 0.23$ and overall frequency of $105, 379$.}
\end{figure}

\Cref{fig:age_gender} splits the Rekognition annotations by imputed gender-age group and displays the pass rate and frequency associated with each group. The age distribution concentrates around ages \texttt{20-29} for the raw and filtered datasets. Controlling for age, we observe higher representation of images in the \texttt{Female}-imputed groups for young ages compared to that of the corresponding \texttt{Male}-imputed groups. These \texttt{Female}-imputed groups are also more likely to pass the CLIP filter, which again reinforces the correlation between representation and pass rate. This trend then reverses for ages above $20$ --- for example, there are more images from the \texttt{Male},\texttt{60-69}-imputed group as well as a higher pass rate compared to images from the \texttt{Female},\texttt{60-69}-imputed group. This finding again supports the notion of exclusion amplification in multimodal filtering: an imputed gender-age group with higher frequency in the raw dataset is roughly associated with a higher pass rate.

\paragraph{Gender and race annotation method:} Given the established biases and shortcomings of pre-trained third-party attribute predictors \citep{gendershades}, in addition to using Amazon Rekognition, we introduce another technique to complement the previously presented results. We follow methodology very similar to \citet{bianchi_easily_2023} and \citet{luccioni2024stable}, which uses embedding clustering techniques on a reference database (in our case, the Chicago Face Database using their self-reported gender and race annotations \citep{cfd}) in order to assess a model's internal representation of gender and race. We defer to \Cref{sec:appx:methods:clip_knn} the additional details about the implementation and validation of this method in comparison to Amazon Rekognition. This method requires face box annotations, so we apply this method to the same set of $11,000$ images detected by Rekognition as containing a single face. Going forward, we refer to this technique as CLIP representation-based kNN, or \textit{CLIP kNN} for short, and we follow the same annotation group names as used in the Chicago Face Database. 

At a high level, we note that CLIP kNN is not equivalent to a demographic group predictor, but rather a way to map an image to the space where CLIP encodes demographic information according to the embeddings of the Chicago Face Database. Therefore, we interpret annotations from this CLIP kNN technique as how the CLIP model internally associates images with self-reported demographic information. Because we aim to assess how CLIP acts as a filter, this allows us to group embeddings from CommonPool and evaluate the impact of filtering by each associated group.

\subsubsection{The CLIP kNN technique confirms the Rekognition findings on imputed gender, where images from the \texttt{Female}-imputed group are less represented in the raw dataset and excluded at slightly higher rates.}
\label{sec:clip_knn_gender}

We confirm similar trends of exclusion amplification by gender on the same subsample, as shown in \Cref{fig:race}. For the \texttt{Female}-imputed group, $608$ out of $2,444$ samples pass the CLIP filter for a pass rate of $0.25$. For the \texttt{Male}-imputed group, $847$ out of $3070$ samples pass the CLIP filter for a pass rate of $0.28$.

\subsubsection{There are differences in how the CLIP filter treats the CLIP kNN clusters by race, and \texttt{White}-imputed images consist of a majority of these images in both the raw and filtered datasets.}

\begin{figure}
    \begin{center}
    \includegraphics[width=0.48\linewidth]{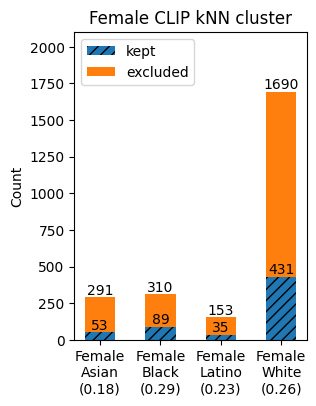}
    \includegraphics[width=0.48\linewidth]{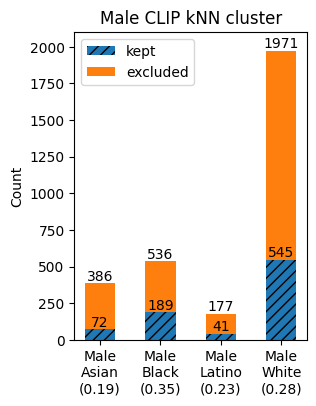}
    \end{center}
    \caption{Frequency split by CLIP kNN according to race annotations with pass rates included in parentheses. A majority of the images in both the filtered and raw datasets belong to the \texttt{White}-imputed group.}
    \label{fig:race}
    \Description{A bar graph of frequency of kept and excluded data split by CLIP kNN-grouped gender and race. The left graph shows frequency for the \texttt{Female}-imputed group, the left graph shows frequency for the \texttt{Male}-imputed group. The \texttt{White Female}-imputed group has the highest total frequency of $1690$, while \texttt{White Male}-imputed group has the highest total frequency of $1971$.}
\end{figure}

By the CLIP kNN technique, out of the same subsample of 100 thousand images that are detected by Rekognition as containing exactly one face, \Cref{fig:race} demonstrates that there are pass rate discrepancies between various race-related clusters. \texttt{Asian}-imputed images are filtered out the most, then \texttt{White}, then \texttt{Latino}, then \texttt{Black}. These relative comparisons still hold when split by gender annotations, although there is a substantial pass rate gap between \texttt{Black Male} (0.35) and \texttt{Black Female} CLIP clusters (0.29). We also find highest frequency of \texttt{White}-imputed images (66.8\%) in the raw dataset, which also holds as most frequent in the filtered dataset.

\section{Language and geography analysis}
\label{sec:western}

Examination of filtering by language, country domain, or geolocation reveals that the filtered dataset overrepresents data associated with Western regions when compared to the raw dataset. As a result, we extend prior findings on Western bias in text filtering \citep{aboutme} to the multimodal setting. When performing the same analysis on English text, we find that these discrepancies between Western and non-Western countries still sometimes hold although the disparity is smaller, which we defer to \Cref{appx:results:english}.

\subsection{Language}
\label{sec:western:language}

\paragraph{Method:} We use \texttt{langdetect} library \citep{langdetect} on a random set of $100,000$ samples to determine the language of the text. While prior work demonstrates that language detection predictions may be affected by the text content \citep{aboutme}, this method enables us to assess filtering differences by language at a large scale.

\subsubsection{English is the most frequent language and is included at significantly higher rates compared to other languages.}

We find that the most frequently detected language is English ($35.5\%$ of the raw dataset) and that English also has the highest pass rate ($0.48$). This is to be expected, given that CLIP is trained on English image-text pairs \citep{clip} and therefore would have more valid embedding representations for English text compared to other languages. 

\subsubsection{Non-Western languages are not well-represented in the raw dataset, and filtering amplifies this further.}

\begin{figure}
    \begin{center}
    \includegraphics[width=0.8\linewidth]{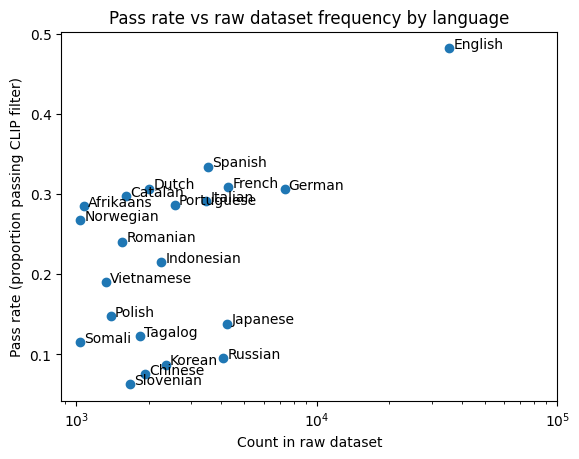}
    \end{center}
    \caption{Pass rate versus raw dataset frequency for common languages with at least 1,000 samples in the raw dataset. Western languages have higher pass rates, and text detected as English comprise of 35,514 samples (out of 100,000 samples) in the raw dataset with a pass rate of 0.48.}
    \label{fig:language}
    \Description{A scatter plot of pass rate vs raw dataset frequency for each language with at least $1,000$ samples. Western languages have higher pass rates, and text detected as English comprise of $35,514$ samples (out of the $100,000$ samples) in the raw dataset with a pass rate of $0.48$. Other languages are more than an order lower in frequency magnitude and have pass rates ranging from $0$ to $0.35$.}
\end{figure}

For languages outside of English, \Cref{fig:language} shows the pass rates of each language (with at least $1,000$ samples) against their frequency in CommonPool. We observe that the next highest pass rates include Western languages like Spanish, Dutch, French, Catalan, German, and Portuguese, potentially due to their similarity to the English language. Furthermore, we demonstrate a similar notion of \textit{exclusion amplification} as the trend from \Cref{sec:demographic_group:image_group}: data from detected languages that have lower representation in the raw dataset are discarded by the filter at higher rates 
(more details, along with a a more comprehensive set of languages, are presented in \Cref{appx:results:language}.

\subsubsection{Our findings reveal that multilingual data obtained after CLIP-filtering skews representation towards Western languages.}

LAION-2B-multi is a subset of LAION-5B containing text in non-English languages, which is also obtained after CLIP-filtering. Researchers argue that this particular dataset is one of the largest multilingual datasets to date and therefore can fuel new research on ``low-resource'' languages \citep{laion5b}. Our findings of Western language bias in CLIP-filtering complicate LAION-2B-multi's contribution --- filtering with a CLIP model trained on English text overrepresents text written in Western languages and does not result in a uniform representation of the distribution of multilingual text on the internet.

\subsection{Country domain}
\label{sec:western:country}

\paragraph{Method: } We examine rates of filtering across country domains as a rough proxy to assess how data relating to certain geographic regions are treated by the CLIP filter. For this analysis, we use all 12.8 million samples from the \texttt{small} version of CommonPool and examine base domains with at least $10,000$ samples from the raw dataset.

\subsubsection{Country domains from non-Western regions are excluded at higher rates than country domains from Western regions.}

\begin{figure}
    \centering
    \includegraphics[width=\linewidth]{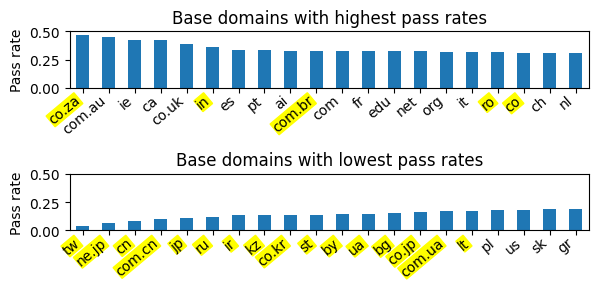}
    \caption{Base domains with at least 10,000 samples sorted by the highest (top) and lowest pass rates (bottom). Highlighted base domains associate with countries not considered in the Rich West \citep{richwest}. We see that most of the base domains with lowest pass rates correspond to non-Western regions.}
    \label{fig:base_domain}
    \Description{Bar graph of the 20 country domains with the highest pass rates on the top and a bar graph the 20 country domains with the lowest pass rates on the bottom. A higher pass rate represents a higher proportion included in the resulting CLIP-filtered dataset. Highlighted country domains associate with countries not considered in the Rich West \citep{richwest}. We see that most of the base domains with lowest pass rates correspond to non-Western regions.}
\end{figure}

\Cref{fig:base_domain} reinforces the presence of Western bias along the source modality. The base domains of websites with the highest pass rates mainly come from Western countries, while the base domains with the lowest pass rates are predominantly associated with non-Western countries.

\subsection{IP address geolocation}
\label{sec:western:ip_address}

\paragraph{Method: } In addition to country domain, we refer to an IP address geolocation database in order to determine the country related to a source domain. Similar to the approach in \citet{c4}, we use the IP2Location Lite IP-Country Database \citep{ip2location}, which has been demonstrated to have high country-level accuracy \citep{komosny2017location, poese2011ip}. We obtain the IP address set using the python \texttt{socket} library for a random selection of 1 million CommonPool samples, which results in 200 thousand unique websites. IP2Location then successfully identifies the country of 95.7\% of the sampled website domains, which corresponds to 770 thousand CommonPool samples. We recognize that websites can be hosted in various data servers around the world, or based on user location, so IP addresses may not accurately represent the location of a website \citep{livadariu2020accuracy}. As such, we use inferred geolocation to complement the other methods we implement to obtain geographic region.

\subsubsection{The inferred geolocation of most websites come from the United States, and samples with IP address locations from Western countries bypass the filter at higher rates.}

\begin{figure}
    \centering
    \includegraphics[width=\linewidth]{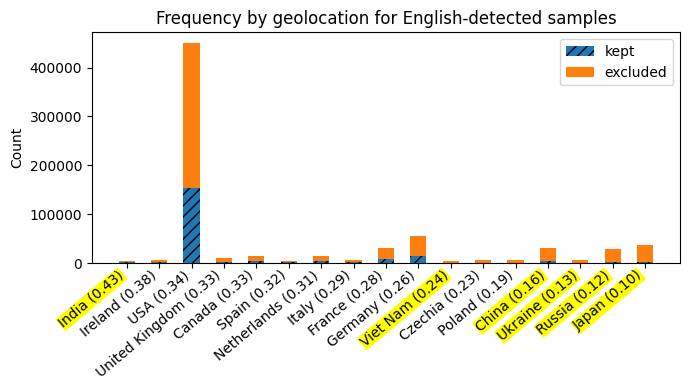}
    \caption{Raw dataset frequency by country of IP-based location sorted in descending order by pass rate (included in parentheses). We display countries with at least 5,000 samples, and most of these non-Western countries (highlighted) have low pass rates.}
    \label{fig:ip_addr}
    \Description{Bar graph of the frequency in the raw dataset by country of inferred geolocation. The USA has the highest frequency of around $450,000$, and all other countries are below $100,000$. Pass rates range from $0.43$ (India) to $0.10$ (Japan).}
\end{figure}

\Cref{fig:ip_addr} demonstrates that for both the raw dataset and filtered dataset, the inferred geolocation of an overwhelming majority of data come from the United States, although this may be due to the origin location used to ping the website. We also find that the IP address locations of countries with higher pass rates correspond to those in Western regions, and that data with inferred geolocations from non-Western regions are filtered out at higher rates.

\section{Source domain analysis}
\label{sec:source}

In this section we highlight results that examine the source domain of DataComp image-text samples and analyze what attributes correspond to being more or less likely to pass the CLIP filter.

\subsection{Websites with high pass rates}
\label{sec:source:website}

\paragraph{Method: } We determine which common websites (with at least $10,000$ samples in the raw dataset) have the highest pass rates. We then manually examine samples from the websites considered ``high-quality'' by CLIP.

\subsubsection{The majority of websites with highest-quality data according to CLIP are stock photo and E-commerce sites.}

Among the top 20 websites with the highest pass rates, \Cref{fig:top_url} shows a large number of websites that upon manual examination we categorize as e-commerce or stock photo platforms. This coincides with results from a recent LAION-Aesthetics v2.6+ audit \citep{baio2022exploring}, which examines frequency in the filtered dataset (rather than pass rates). We note the types of websites that are common from the prior audit but do not have high pass rates, such as user-generated content platforms like Pinterest or Wordpress.

\subsubsection{Some stock photo images that are included in the final dataset are thumbnail images that do not have watermarks.}

Manual inspection of image-text samples from stock photo sites reveals high-resolution images with watermarks and thumbnail images without watermarks, which indicate that these images may have been scraped without the corresponding license for these websites. Shutterstock, for example, requires a license purchase before use of their images \citep{shutterstock}, and it is unclear if distributing links containing these images requires a license.

\subsubsection{Intellectual property from educators intending to be purchased is also scraped as data without compensation.}

Additionally, \textit{Teachers Pay Teachers}, another website with a high pass rate, is an online education content platform, where teachers create their own curriculum to sell to other teachers. These images in DataComp are often worksheets, which therefore disseminate the exact products that are the intellectual property of the curriculum creator \citep{teachers}. These findings have potential implications in the ongoing copyright discussion of training large models \citep{nytimes, copyright, fairuse}, which we expound upon in \Cref{sec:discuss:consequences:copyright}. 

\begin{figure}
    \begin{center}
    \includegraphics[width=\linewidth]{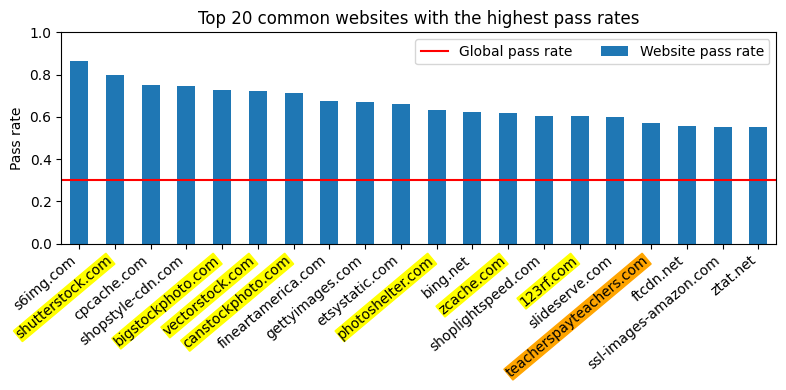}
    \end{center}
    \caption{The top 20 websites with the highest pass rates sorted in descending order. Stock photo websites are highlighted in yellow, and Teachers Pay Teachers, an online marketplace where educators can buy and sell education curriculum, is highlighted in orange. Many of these sites where the majority of their data bypass the CLIP filter are platforms where the image assets themselves are considered copyrighted material intended to be purchased.}
    \label{fig:top_url}
    \Description{Bar graph of the top 25 URL domains with the highest pass rates that range from around $0.85$ to $0.6$. Stock photo websites include shutterstock.com, bigstockphoto.com, vectorstock.com, canstockphoto.com, photoshelter.com, zcache.com, 123rf.com.}
\end{figure}

\subsection{News sites}
\label{sec:source:newsite}

\paragraph{Method: } We look at pass rates of popular news sites \citep{newsite} that each contain at least 200 samples.

\subsubsection{The CLIP filter considers this collection of news sites (especially from the US and UK) as higher quality than average.} 
\begin{figure}
    \centering
    \includegraphics[width=\linewidth]{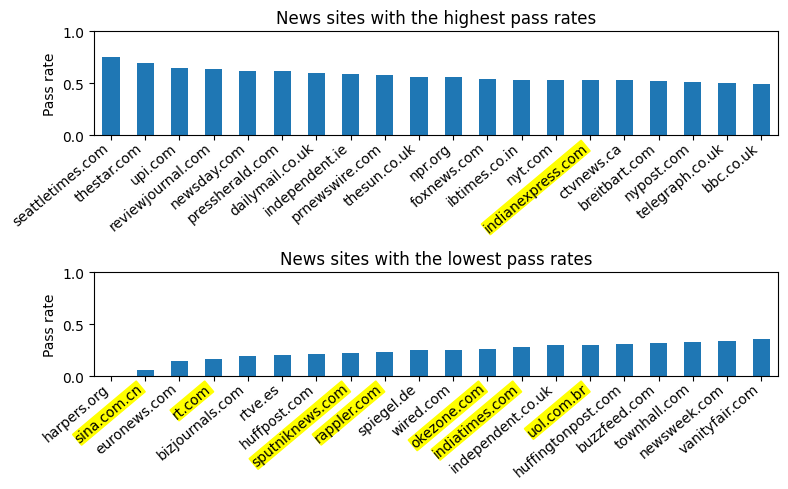}
    \caption{New sites with the highest pass rates (top) and lowest pass rates (bottom). News platforms that correspond to non-Western regions are highlighted in yellow, and we see high non-Western presence among news sites with lowest pass rates.}
    \label{fig:newsite}
    \Description{Bar graph of the top 25 news platforms with the highest pass rates on the top and a bar graph of the bottom 25 news platforms on the bottom. The highest pass rates range from $0.8$ to $0.5$, and the lowest pass rates range from $0$ to $0.3$.}
\end{figure}

We find on average higher CLIP similarity scores for samples from news sites compared to overall samples. However, \Cref{fig:newsite} illustrates that news sites from the United States and the United Kingdom have much higher pass rates compared to news sites from other countries, which reinforces the Western bias highlighted in \Cref{sec:western}. Given the ongoing lawsuit between The New York Times and OpenAI \citep{nytimes}, according to \Cref{fig:newsite}, we note that data from The New York Times has substantially higher pass rates than average.

\subsection{Website category}
\label{sec:source:category}

\begin{figure*}
   \begin{subfigure}{\textwidth}
     \centering
     \includegraphics[width=\linewidth]{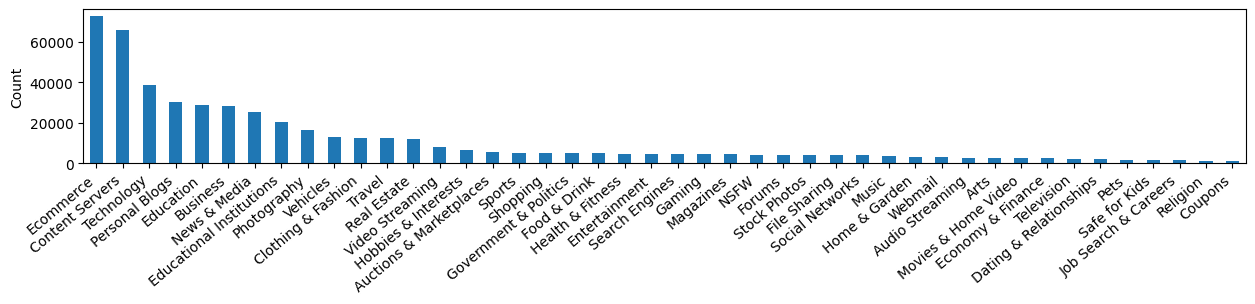}
     \caption{Category frequencies sorted in descending order.}
     \label{fig:cloudflare_raw_count}
   \end{subfigure}
   
   \begin{subfigure}{\textwidth}
     \centering
     \includegraphics[width=\linewidth]{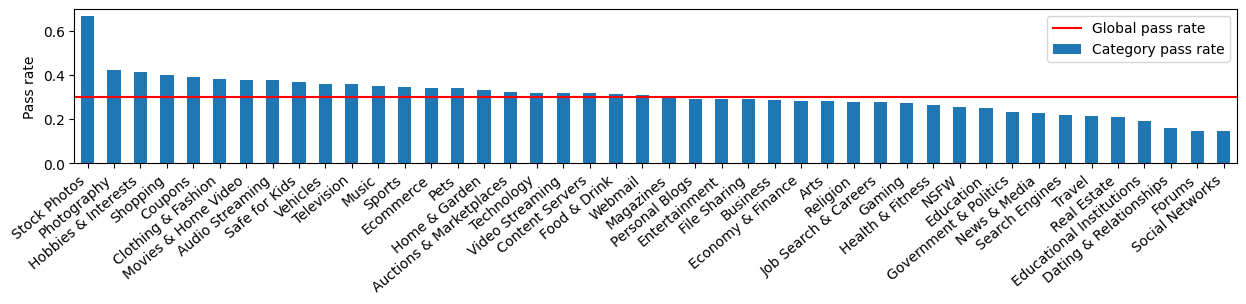}
     \caption{Category pass rates sorted in descending order. The global pass rate (0.3) is plotted in red. }
     \label{fig:cloudflare_filter_ratios}
   \end{subfigure}
   \caption{Website categorization by Cloudflare API \cite{cloudflare} for websites with at least 1000 samples.}
   \Description{Bar graphs of raw count at the top and pass rate at the bottom, grouped by Cloudflare category predictions. Frequency ranges from about $70,000$ (e-commerce) to $1000$ (coupons), and pass rates range from $0.65$ (stock photos) to $0.15$ (social networks).}
\end{figure*}

\paragraph{Method: } Inspired from prior work on text data filtering \citep{aboutme}, we also investigate whether CLIP acts as a topical domain filter at the website level. Because prior work shows that the Cloudflare Domain Intelligence API \citep{cloudflare} has high accuracy across categories \citep{ruth2022world}, we apply this API to categorize a random subset of 100 thousand websites.

The Cloudflare API \citep{cloudflare} returns category predictions for $94,428$ out of the $100,000$ website domains, where a website can correspond to multiple categories. This corresponds to $94$ categories across roughly $400,000$ samples in CommonPool, and we examine categories with at least $1,000$ samples. Since some of these categories overlap in meaning or are confusingly defined, we merge the categories in the same manner as \citet{ruth2022world} and extend it to new Cloudflare categories, which we defer to \Cref{appx:methods:cloudflare}.

\subsubsection{E-commerce, image-hosting, and user-generated websites are among the most popular Cloudflare-detected categories in the raw dataset.}

\Cref{fig:cloudflare_raw_count} reveals a breakdown by website category in the raw dataset. The most common categories are \texttt{E-commerce}, \texttt{Content Servers} (sites that host static images), \texttt{Technology},\newline \texttt{Personal Blogs} (user-generated content), and \texttt{Education}. This again matches prior analysis of a subset of LAION \citep{baio2022exploring} with a high representation of shopping-related and user-generated websites. In our case, however, we find there are relatively fewer samples from the \texttt{Stock Photos} category compared to more popular categories.

\subsubsection{Data from stock photo sites are more likely to be included by the CLIP filter.}

The most common categories are not necessarily included at higher rates, as \Cref{fig:cloudflare_filter_ratios} shows that the \texttt{Stock Photos} category has the substantially highest pass rate ($20$ percentage points higher than the next category, \texttt{Photograph}). We manually verify that the websites that fall under this category consist of mainly stock photo sites, which confirms the findings by website domain from \Cref{fig:top_url}. In contrast to \Cref{sec:source:newsite}, we find that the \texttt{News \& Media} category has a lower pass rate than average, possibly because this category contains a broader range of websites than the popular news dataset we use.

\subsubsection{We find presence of sexually-explicit text content from NSFW-categorized websites that passes both the NSFW and CLIP filters.}
\label{sec:source:category:nsfw}

Figures \ref{fig:cloudflare_raw_count} and \ref{fig:cloudflare_filter_ratios} reveal that some websites are categorized by Cloudflare as \texttt{NSFW}, which includes content relating to pornography, nudity, extremism, and violence. This corresponds to $4,104$ samples before CLIP-filtering (out of the $400,000$ CommonPool samples with websites categorized) and $1,038$ kept after CLIP-filtering. Upon this finding, because the image-text samples themselves may not contain NSFW, we manually examine the text of these samples to flag sexually-explicit data. Of the $4,000$ or so text samples, we find $211$ samples with sexually-explicit text content in CommonPool, and of these, $11$ samples that pass the CLIP filter. Five sexually-explicit text samples contain words like ``teen'' and ``schoolgirl,''  and we have chosen not to visually examine these images.

It is not surprising that sexually-explicit text passes through the CLIP filter as CLIP is not trained to detect NSFW-related content. However, in the formation of CommonPool, an NSFW text filter was applied to the initial web dump in order to remove sexually-explicit text \citep{datacomp}. Given we only categorize the websites of $400,000$ random samples due to API limits, we find that $1.03\%$ of samples of this small subset come from NSFW-categorized websites. Extending to all $12.8$ billion samples of CommonPool, at a $95\%$ confidence interval of this pass rate estimate, the number of samples that come from websites categorized as NSFW by Cloudflare is between $129$ million and $130$ million, of which $32$ to $34$ million samples would pass the CLIP filter. In the case of the proportion of sexually-explicit text content in our subsample, at a $95\%$ confidence interval, the number of samples in CommonPool with sexually-explicit text is between $6.3$ million and $7.2$ million, of which $250,000$ to $450,000$ samples would pass the CLIP filter. We also note that these numbers are a substantial lower bound as we examine only English text, ignore image content, and consider samples solely from websites categorized as NSFW by Cloudflare --- more investigation is needed here, but given that NSFW-filtering is outside of our original evaluation scope, we leave follow-up analysis to future work.

\subsection{Origin date}
\label{sec:source:date}

\paragraph{Method: } To track the creation date of a URL address, we apply the Wayback Machine API \citep{wayback} to a subsample of 1 million image URL addresses to track the earliest-indexed date in the Internet Archive. We acknowledge that many webpages may not have snapshots on the Internet Archive and that the earliest-indexed date is only an upper bound for when the webpage was actually created.

\subsubsection{Most data comes from the last five years, according to earliest-indexed Internet Archive dates.}
The Wayback Machine API \citep{wayback} responses indicate that the Internet Archive does not record snapshots for a majority of the exact URL addresses in our subsample --- we only track index dates for 22.7\% of the one million CommonPool samples. Of those tracked, we find that these samples overwhelmingly come from sources that were first indexed within the last ten years. We also find that pass rates tend to be somewhat higher the more recent the year. We present the graphs and details in \Cref{appx:results:year}.

\section{Discussion}
\label{sec:discuss}

Our analysis demonstrates that CLIP-filtering on the DataComp CommonPool dataset removes data nonuniformly when broken down by demographic group imputation, geographic region, or source domain. First, we find that the CLIP filter disproportionately excludes data relating to various proxies for demographic groups which include LGBTQ+ identities, women, and non-Western regions. Second, these filter discrepancies often amplify representation biases present in the raw dataset, where underrepresented groups are filtered out at higher rates compared to overrepresented groups. Third, we find that data from sites where the images themselves are the intellectual property (e.g., stock photos and educational curriculum) are considered high-quality by CLIP and kept at higher rates. Finally, the presence of NSFW samples after an initial cursory search of a small subset of CommonPool reveals that the combination of applying NSFW and CLIP filters fails to catch all instances of sexually-explicit content.

In this section, we elaborate on how the above findings imply that CLIP-filtering and similar filtering practices may result in societal harms and potential legal implications. Using the CLIP filter as a case study, we then dive into several key assumptions and failures of existing data cleaning methods when they are used to build large image-text models. For instance, we discuss how the CLIP filter may be more likely to pass samples similar to its own training data in \Cref{sec:discuss:assumptions:alignment}. Finally, we use our analysis to inform recommendations for designing, implementing, and evaluating data filtering techniques and data curation methods.

\subsection{Societal consequences of data filtering}
\label{sec:discuss:consequences}

We highlight three aspects of CLIP-filtering CommonPool data that have societal ramifications: the amplification of demographic bias, the inclusion of unmoderated inappropriate content, and the upweighting of websites that host licensed or copyrighted images. Altogether these dimensions illustrate an example of how existing data filtering practices have failed to mitigate certain downstream harms and, hence, that stochastic neural-network-based filtering may not be the end-all-be-all technique to capture \textit{all} ``impurities'' in web-scraped data.

\subsubsection{Disproportionate exclusion from filtering may worsen downstream models.}
\label{sec:discuss:consequences:exclusion}

As mentioned in \Cref{sec:approach:limits}, we cannot necessarily jump to conclusions that all else held equal, varying the demographic group will change the filter likelihood, and we leave this type of causal investigation~\citep{counterfactual} for our future work. It is, for example, possible that more data relating to a certain group can come from websites with established alt-text practices, which would change filtering rates even without using a biased model like CLIP. At the same time, however, the finding of correlations between imputed demographic groups and filtering rates is important absent this causal measurement. For CommonPool, the CLIP filter modifies the original demographic distribution and thus does not result in a uniform sampling of web-scraped data. \textit{Our analysis in \Cref{sec:demographic_group:identity} shows that CLIP-filtering can create training datasets more unrepresentative of certain populations and containing higher proportions of stereotypes than the unfiltered datasets.}

Because we find that the CLIP filter excludes data relating to certain imputed demographic subgroups, a lack of data from these subgroups can therefore impact downstream models trained on CLIP-filtered datasets. Prior work demonstrates that diverse representation in training is necessary to ensure model accuracy for minority subgroups \citep{rolf2021representation, chen2018my}. The exclusion amplification trend we find can further compound these model inequities: web-scraped data already does not encompass diverse views \citep{suchin}, yet CLIP-filtering may worsen the amount of data homogeneity in the resulting dataset. As a consequence of this form of representational harm \citep{nlpbias}, machine learning models trained on data obtained via biased filtering can underperform even more for certain marginalized groups. 

In addition, we observe that the CLIP filter is more likely to include stereotypical associations between mentions of certain words and genders (\Cref{sec:demographic_group:identity:word_stereotypes}). This indicates that CLIP-filtering may propagate and thus embed discriminatory associations into downstream datasets. We hypothesize this bias can lead to generative models defaulting to stereotypes of marginalized groups~\cite{wired_lgbtq, verge_bias}.

\subsubsection{Continued presence of lightly or unmoderated sexually-explicit material raises CSAM and NCII concerns}
\label{sec:discuss:consequences:nsfw}

Based on the findings in \Cref{sec:source:category:nsfw}, out of all 12.8 billion samples of CommonPool, we estimate approximately 130 million samples from websites categorized as NSFW by Cloudflare make it past NSFW-filtering into CommonPool, and approximately 33 million of those samples pass CLIP-filtering. While additional investigation is needed to assess CommonPool's image content, we observe that sexually-explicit text samples are not caught by DataComp's NSFW-filtering step. Given that the direct predecessor to CommonPool, LAION-5B, was found to contain thousands of images of child sexual abuse material (CSAM) and that some of these images were not tagged by LAION's NSFW classifier ~\citep{thiel_identifying_nodate}, we cannot assume that both CommonPool and CLIP-filtered CommonPool do not contain CSAM, even with the prior application of an NSFW filter.

In addition to the prevalence of nonconsensual intimate imagery (NCII) on the web \citep{langlois2017economies}, we repeat the argument \citet{birhane} made against LAION-400M in 2021: neural network-based filtering cannot provide guarantees of removing all NCII and CSAM. Methods other than neural-network-based filtering are likely to be required in order to remove NCII and CSAM from web-scraped image datasets. Merely changing or updating filters, as has been done since 2021, is insufficient, as our analysis shows. These ongoing, well-documented failures are causing real, severe harms at scale: text-to-image models trained on these ``filtered'' datasets are being used to generate NCII of countless celebrities and random, regular people~\citep{verge_taylor, aoc} as well as CSAM~\citep{iwf, laion_csam}, with teenage girls being targeted in particular~\citep{nytimes_deepfake}. Because of the AI field's rapid embrace of these datasets and models despite widespread documentation of the deep NCII and CSAM harms they are causing, we urge AI industry, academia, and other players to prioritize these issues.

\subsubsection{Certain types of copyrighted data are considered more ``valuable'' in the machine learning pipeline.}
\label{sec:discuss:consequences:copyright}

There has been extensive discussion on the copyright implications of generative models trained on large-scale datasets scraped from the web. In the recent lawsuit from The New York Times against OpenAI, for instance, the plaintiff has argued that The New York Times articles are considered more ``valuable'' during the training process \citep{nytimes}, where ``value'' here refers to the amount of content contributed to training a model. OpenAI in response has claimed that these articles are just ``a tiny slice of overall training data'' \citep{openai_response}. Our analysis in \Cref{sec:source:newsite} provides additional evidence that data from certain websites, including The New York Times, are implicitly upweighted after filtering, which implies that systems trained through this kind of filtering do in fact place more value on certain types of copyrighted works, although it is unknown if OpenAI implements a similar filtering mechanism in constructing its training data.

In Sections \ref{sec:source:website} and \ref{sec:source:newsite}, we show that certain kinds of typically copyrighted work pass the CLIP filter at significantly higher rates --- namely stock photo sites, some news sites, and an education curriculum marketplace --- all of which rely on their image and textual content as part of their business model. We also find presence of stock photo thumbnail images without watermarks, despite no known licensing agreement from the dataset curators to distribute these photos \cite{shutterstock}. This finding indicates that data from these platforms are on average considered ``high-quality'' sources by CLIP, which means that the CLIP filter overrepresents and implicitly targets data from these copyrighted websites.

In order to make a case for copyright infringement in the United States, according to the fair use doctrine, the defendant must demonstrate that use of copyrighted material follows four factors \citep{law}. According to legal analysis of fair use in the context of generative machine learning models, the fourth factor of fair use, which examines the potential market effect, draws the most attention from legal scholars \citep{copyright, fairuse}. \citet{alhadeff2024limits}, for example, argue that the fourth factor weighs against a finding of fair use because the output generative AI tools may be a substitution for market harm.

Text-to-image models that train on this data without purchasing licenses can therefore replicate these images which can lead to potential business losses. Users, for instance, may no longer license from stock photo sites if they can produce images on their own, or educators may generate worksheets rather than buying from curriculum developers. Our findings thus provide more context on the potential market harm as described in fair use copyright doctrine \citep{alhadeff2024limits, copyright, fairuse}, especially if these data creators are not compensated. Dependent on additional legal analysis, this raises questions on whether plaintiffs in copyright infringement lawsuits, such as The New York Times \citep{nytimes}, may be able to seek higher damages as filtering inadvertently deems their data as higher value.

\subsection{Assumptions in data filtering practices}
\label{sec:discuss:assumptions}

To understand how the above issues were overlooked in the design of CLIP-filtering, we examine how prior work justified certain filtering choices. This enables us to form takeaways about OpenAI CLIP as a filtering model and about data filtering more generally.

\subsubsection{OpenAI CLIP is not intended to assess image-text alignment.}
\label{sec:discuss:assumptions:alignment}

In our work, we demonstrate instances of stereotypes perpetuated by the CLIP filter when examining common words or intersections of demographic dimensions (Sections \ref{sec:demographic_group:identity:intersections} and \ref{sec:demographic_group:identity:word_stereotypes}), as well as instances of Western bias (\Cref{sec:western}). These trends support the notion that the OpenAI CLIP model is not intended to assess \textit{image-text alignment}, as pointed out in prior work on image captioning CLIP evaluations \citep{qiu_gender_2023} and on CLIP classification bias \citep{agarwal_evaluating_2021}. Especially because OpenAI CLIP was never designed to be a data filter, much less a filter used to build deployed models \cite{modelcard}, we strongly caution against using CLIP as a filter to build future training datasets.

It is entirely plausible that OpenAI CLIP is trained on a distribution of data similar to the filtered dataset makeup, which means the CLIP filter may implicitly attempt to replicate its own training dataset. Our findings on exclusion amplification confirm this notion, as well-represented classes of data seem to be included at higher rates. However, none of this can be confirmed given that the training data for OpenAI CLIP has not been released, nor have the data filtering steps to obtain said training data been formally described~\citep{clip}. In other words, the CLIP model is generated by a black-box pipeline, trained on an unknown dataset. This results in a black-box filtering model whose behavior is difficult to predict without prohibitively many interactions with it. The main advantage of CLIP-filtering here is automation, which embodies the existing notion of how ``scale beats noise'' as characterized in \citet{birhane}.

\subsubsection{Data filters will always encode ideology as to what is considered ``high-quality.''}
\label{sec:discuss:assumptions:quality}

Our study corroborates past work on text filtering bias \citep{suchin, aboutme, c4} in which image-text filtering also encodes ideology of what sociodemographic identities are associated with ``high-quality'' data. In the case of CLIP, we determine that the CLIP filter considers data relating to overrepresented groups as supposedly high-quality. Regardless of what a data filter lets through, the act of discarding data involves judgement calls of what is considered ``good'' or ``bad'' data.

Prior works on data filtering focus on improving the quality of the resulting dataset \citep{gpt3, wenzek2019ccnet}, and there has been subsequent work examining the tradeoffs between quality and quantity \citep{nguyen2022quality, goyal2024scaling}. This notion of ``quality,'' however, is not well-defined and is often treated as an inherent component of the data that must somehow be discovered. As a result, filtering is depicted as a passive act, where issues of the data are blamed on the state of the world. In this manner, machine learning practitioners and researchers are able to avoid the responsibility of addressing societal inequities or problematic content \citep{dafoe2015technological, hanna2020against}.

Investigating the design of CLIP-filtering, we find assumptions ingrained into the objectives of the filter. The LAION-400M developers optimize for a cosine-similarity threshold to improve the performance on the downstream CLIP model \citep{laion}, but this seemingly neutral objective is encoded with implicit priorities. ``Performance'' in the DataComp benchmark refers to improvement on a specific suite of evaluation image classification and retrieval tasks, which mainly measure object classification or distribution shift \citep{datacomp, fang_data_nodate}. \citet{values} state that performance is often considered ``intrinsic,'' yet current measures of what constitutes as success for data filtering ignore societally-relevant concepts like fairness, toxicity generation, or privacy preservation, all of which are risks of text-to-image models \citep{bird2023typology}. CLIP-filtering is argued to be effective, but only because it aligns with a specific notion of performance, in which justification behind the selection of these metrics is unstated.

\subsection{Recommendations}
\label{sec:discuss:recommendations}

After questioning the assumptions of current data filtering practices, we subsequently form several recommendations. While most of these recommendations apply to data filtering methods specifically, we argue that that data filters cannot be treated as a panacea \citep{birhane}. As such, we extend some of our recommendations to the data curation process more generally.

\subsubsection{Intentionally consider filtering criteria to account for diversity}
\label{sec:discuss:recommendations:diversity}

All these lines of inquiry demonstrate the need to build filters that are explicitly designed to include representative and non-stereotypical data from marginalized groups. Like prior works on data filtering biases \citep{suchin, c4}, we also recognize the need to incorporate inclusive data collection practices \citep{parrots, jo2020lessons}, especially concentrating on ethical practices for human-centric data \citep{andrews2024ethical, scheuerman2021datasets}. If these large-scale web-scraped image-text datasets continue to be built, it becomes necessary to investigate how to build filters that do not perpetuate representational harms. Creating fairer filters that allow for more diverse data is an open question we leave for future work, and extending the data filtering network framework from \citet{fang_data_nodate} to downstream model bias evaluations can be a promising first step.

\subsubsection{Provide justification for data filter design}
\label{sec:discuss:recommendations:justification}

In addition to intentional data curation, we also argue that it is important to justify and unpack assumptions in filtering design choices. Given the vague definition of ``performance'' in machine learning literature \citep{values}, we recommend that work proposing new data filtering methods critically examine choices in the evaluation and optimization of a filtering model. Filtering is bound to make judgement calls on data, so one should understand the different impacts of various design decisions and provide justification as to what should be considered ``quality'' data. Because models trained on filtered data may be deployed in high-stakes settings, we also encourage considering user-centered design practices to build data filters \citep{vredenburg2002survey}.

\subsubsection{Report and evaluate data filters}
\label{sec:discuss:recommendations:eval}

We recognize the importance of documentation for filtering techniques. Academic benchmarks like DataComp's filtering track and the open release of CommonPool \citep{datacomp} enable our evaluation of image-text filtering bias. Many commercially-deployed models do not release their data collection processes, much less their filtering methods, and it is unclear how many pre-trained filter models like CLIP are applied to large-scale datasets. We believe that making datasets like LAION and DataComp publicly available for independent audits is vital to the consequences of data filtering practices \citep{den}. However, public release must include stronger safeguards against misuse. CLIP and LAION state they are not meant for real-world production contexts or applications, yet they have been used to create products with millions of users \citep{stablediffusion, midjourney}. DataComp CommonPool is not intended for production-ready products, but grants permission for anyone to train models on their datasets, whether deployed or not \citep{datacomp}. We encourage dataset curators to use licenses that restrict against misuse, such as the RAIL license \citep{rail}.

Moving forward, we argue that to even begin to build fairer filters, one must evaluate them first. We recommend future work on proposed filters to evaluate their filter models for bias building upon our approach. Our analysis reveals that the filtering step is a stage where harmful stereotypes can be injected and amplified, especially as filters can be applied to create multiple datasets. Therefore, to build multimodal models that exhibit less biased and problematic behavior, one should assess the strengths and limitations of a proposed filtering approach to account for long-term implementations on future datasets. We also recognize that it is important to conduct bias evaluations in each step in the broader machine learning pipeline \citep{yang_fairness-aware_2020} and to continue to do so as the pipeline evolves.

\subsubsection{Account for prevention and detection of problematic content, as filters are insufficient to create legal or ethical datasets.}
\label{sec:discuss:recommendations:defense}

Our work demonstrates that existing data filters do not capture all instances of problematic content. Existing NSFW or toxicity detection mechanisms are often flawed \citep{c4, toxic-bias}, which indicates more research is necessary in this area. Nonetheless, additional prevention and response mechanisms are needed because filtering cannot guarantee the removal of all undesirable content \citep{lee2020detecting}. Drawing from a defense-in-depth security approach \citep{mughal2018art}, we recommend that researchers and dataset curators combine multiple mechanisms, which includes providing methods of recourse to take down problematic data and update downstream models accordingly \citep{nguyen2022survey}. In \Cref{sec:appx:interventions}, we highlight some potential interventions to existing datasets that contain NSFW content and discuss their implications.

Moreover, our findings corroborate the need to incorporate third-party audits into the machine learning development pipeline. For instance, our audit reveals that removing samples from domains categorized NSFW by Cloudflare is a simple practical method to reduce problematic content from one avenue, although we acknowledge that this method is not a comprehensive remedy. The presence of NCII and CSAM in even filtered web-scale datasets, along with the inability of automated tools to provide rigorous guarantees of their removal, poses a fundamental challenge to web-scale AI.

\section{Conclusion}

In this work, we audit the CLIP-filtering step in the DataComp and LAION pipelines, where to construct ``high-quality'' large-scale training datasets from web-scraped data, an existing pre-trained CLIP model is used to determine image-text alignment. We find that current forms of image-text filtering, similar to prior work \citep{suchin, aboutme, c4}, embed societal judgements in determining what kinds of data should be discarded. Specifically, we find that content relating to marginalized identities or non-Western regions are more likely to be left out of the final training dataset. Moreover, we demonstrate examples of \textit{exclusion amplification} -- data from certain imputed demographic groups already underrepresented in the unfiltered dataset are filtered out at higher rates. We find that copyrighted data from certain types of websites are judged to be higher quality and that NSFW filters fail to remove large quantities of sexually-explicit text, raising CSAM and NCII concerns. These findings raise new issues on the downstream societal effects of filtering methods within the broader machine learning pipeline, especially as large multimodal models become increasingly deployed in high-stakes settings \citep{ali_evaluating_2023, booth2021bias}.

\begin{acks}

We would like to thank Max Del Real, Kentrell Owens, Miranda Wei, and Christina Yeung for their helpful feedback, as well as Lucy Li for the occupation dataset. We are grateful to the Cloudflare Domain Intelligence team, as well as Kimberly Ruth and Sudheesh Singanamalla, for their help with the Cloudflare API. We also thank Alex Fang, Thao Nguyen, Mitchell Wortsman, Pang Wei Koh, and Ludwig Schmidt for our initial discussions on data filtering.

The first author is supported by the NSF Graduate Research Fellowship Program. This work was supported in part by U.S.\ National Science Foundation awards CNS-2205171 and CCF-2045402, the Carnegie Bosch Postdoctoral Fellowship, and a grant from the Simons Foundation.

\end{acks}

\bibliographystyle{ACM-Reference-Format}
\bibliography{ref}

\clearpage

\appendix
\onecolumn

\section{Methodology details}
\label{sec:appx:methods}

In this section, we highlight additional details on the demographic group imputation techniques and website categorization method.

\subsection{Gender keywords}
\label{sec:appx:methods:gender_keywords}

\Cref{tab:gender_keywords} describes the keywords relating to mentions of women and men in the text samples, in order to perform analysis in pass rates by common words in \Cref{sec:demographic_group:identity:word_stereotypes}.

\begin{table}[ht]
\caption{List of regular expressions relating to gender keywords for women and men.}
\label{tab:gender_keywords}
\begin{center}
\begin{tabular}{lll}
\hline
Women-related &
Man-related \\
\hline
\texttt{wom[ae]n} &
\texttt{m[ae]n} \\
\texttt{females?} &
\texttt{males?} \\
\texttt{(she|her|hers)} &
\texttt{(he|him|his)}
\\ \hline
\end{tabular}
\end{center}
\end{table}

\subsection{CLIP kNN}
\label{sec:appx:methods:clip_knn}

The CLIP representation-based kNN method (used in \Cref{sec:clip_knn_gender}) is an audit technique to evaluate CLIP associations with demographic markers like gender and race without ground-truth demographic annotations. Based on the method from \citet{bianchi_easily_2023}, we use the Chicago Face Database \citep{cfd} as a reference database to cluster CLIP embeddings of CommonPool images. The steps are defined as follows:

\begin{enumerate}
    \item We obtain the CLIP embeddings for images in the Chicago Face Database, or CFD, which contain self-reported gender and race attributes corresponding to the images of 597 unique individuals \citep{cfd}.
    \item We then consider image samples from our CommonPool subsample that have a single face detected by Rekognition \citep{rekog}. We crop these images to the bounding box.
    \item Given a facial image, we extract the CLIP embedding and run $k$-nearest neighbors with respect to the CFD embeddings. Note that we use separate kNN classifiers for gender and race. We select $k = 7$ for gender and $k = 5$ for race to maximize validation accuracy on held-out CFD images, and use Minkowski distance as the distance metric.
    \item The distribution of the nearest neighbors becomes the probability scores of the gender or race of the DataComp image, and the argmax is the final group annotation.
\end{enumerate}

\begin{figure}[ht]
   \begin{subfigure}{0.45\linewidth}
     \centering
     \includegraphics[width=\linewidth]{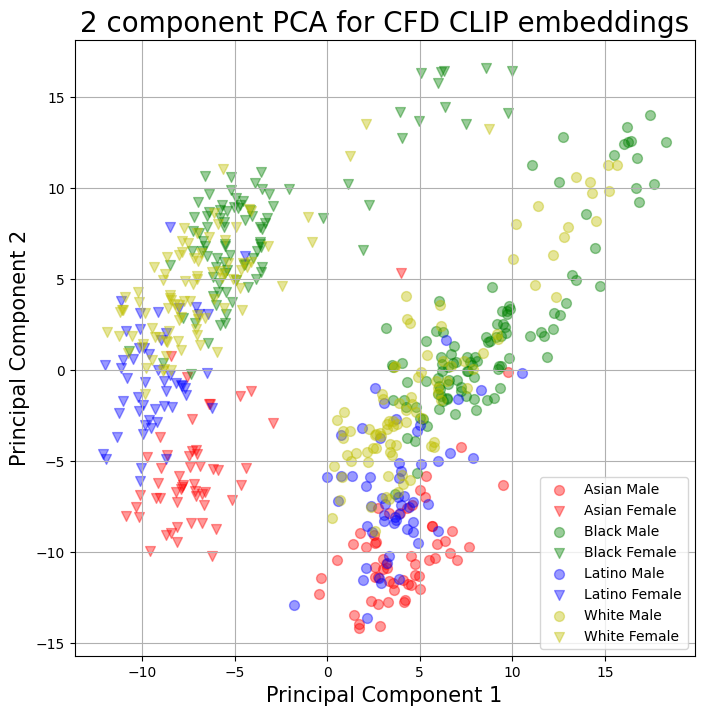}
     \caption{PCA}
   \end{subfigure}
   \hfill
   \begin{subfigure}{0.45\linewidth}
     \centering
     \includegraphics[width=\linewidth]{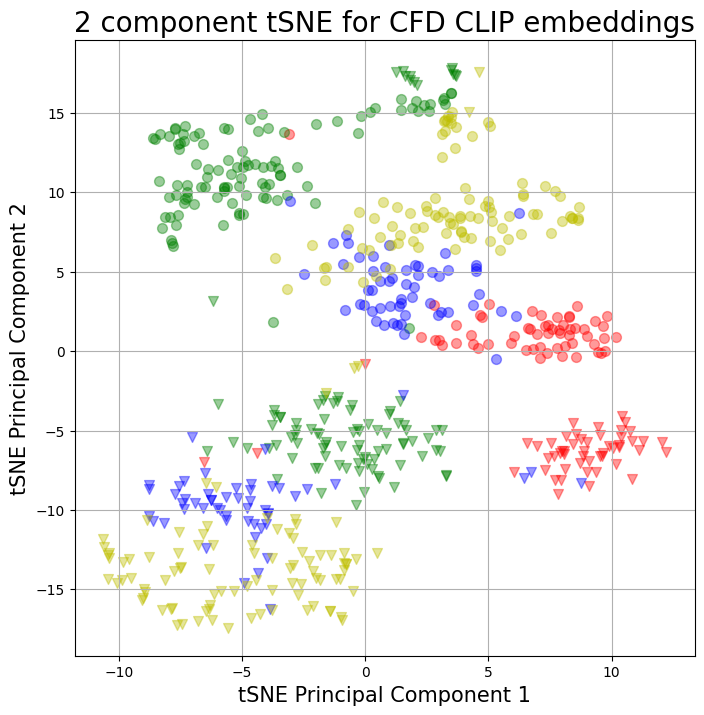}
     \caption{tSNE analysis}
   \end{subfigure}
   \caption{Visualizations of CLIP embeddings of Chicago Face Database images \citep{cfd}, grouped by gender-race annotations. We observe clear clustering by gender and race, although within the same group, embeddings fall into multiple clusters.}
   \label{fig:cfd_viz}
   \Description{Two scatter plots of two-dimensional visualizations of CLIP embeddings of Chicago Face Database images via PCA on the left and tSNE on the right.}
\end{figure}

This process exactly follows the method in \citet{bianchi_easily_2023} with CFD as the reference dataset, except for steps 3 and 4, in which their work calculates the average of all embeddings grouped by gender or race in order to obtain an ``archetypal vector representation'' for a demographic category. Then, for a given test image embedding they select the group with the closest average embedding in cosine distance.

Initially, we find that the CFD embeddings are clustered by demographic group as shown in \Cref{fig:cfd_viz}. However, within the same gender and race group, there are multiple clusters at different regions in the embedding space. This indicates that averaging may not be as effective as a clustering-based classification method, which motivates our use of k-nearest neighbors instead.

\subsubsection{Validation}

On $160$ randomly held-out CFD images, the gender annotation method obtains a validation accuracy rate of $100\%$ (across all gender-race groups). The race annotation method obtains a validation accuracy of $90.6\%$, but we note that the \texttt{Latino Female} group obtains the lowest accuracy of $65.0\%$.

\begin{figure}[ht]
   \begin{subfigure}{0.49\linewidth}
     \centering
     \includegraphics[width=\linewidth]{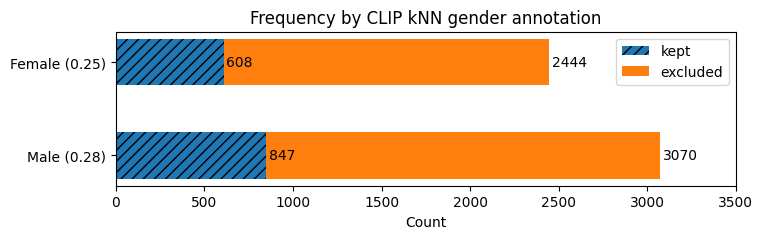}
     \caption{CLIP-kNN}
   \end{subfigure}
   \hfill
   \begin{subfigure}{0.49\linewidth}
     \centering
     \includegraphics[width=\linewidth]{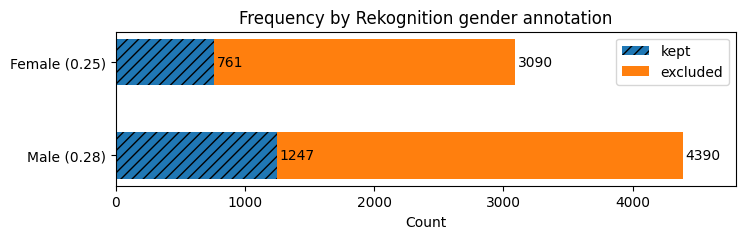}
     \caption{Rekognition}
   \end{subfigure}
   \caption{Frequency by imputed gender for both CLIP-kNN and Rekognition methods applied to a set of 11,000 images that Rekognition detects as containing a face. We see similar trends of exclusion amplification where there are more images from the \texttt{Male}-imputed group than images from the \texttt{Female}-imputed group. Moreover, the pass rate for the \texttt{Male}-imputed group is higher than that of the \texttt{Female}-imputed group. We note that CLIP-kNN annotates fewer total images than Rekognition due to the confidence threshold.}
   \label{fig:gender}
   \Description{Bar graphs of frequency by imputed gender for the CLIP-kNN method on the left and the Rekognition method on the right. The CLIP-kNN method annotates $2444$ total samples as Female with a pass rate of $0.25$ and $3070$ samples as Male with a pass rate of $0.28$. The Rekognition method annotates $3090$ total samples as Female with a pass rate of $0.25$ and $4390$ samples as Male with a pass rate of $0.28$.}
\end{figure}

We find that the CLIP kNN method closely agrees with the gender predictions of Amazon Rekognition when applied to the same subset of $11,000$ images that Rekognition detects as containing a single face. With a confidence threshold set to include only samples where all $k$-nearest neighbors (in CFD) have agreeing ground-truth gender annotations, we find that $95.4\%$ of the $5,514$ CLIP kNN annotations match the Rekognition annotations. Of the $254$ samples that disagree, manual examination reveals that the CLIP kNN annotation more often aligns with human annotation. Again, we note that the CLIP kNN method is not meant to be interpreted as a perfect prediction of gender, especially as gender is not a visual construct \citep{gender}, but rather to assess CLIP's internalized encoding of the gender attribute. This technique allows us to evaluate whether the CLIP filter treats these internal associations differently. \Cref{fig:gender} show that both methods support trends of \textit{exclusion amplification} by imputed gender.

\subsubsection{Limitations}

We recognize that the kNN annotation method relies on face detection boxes as input, which requires an accurate face detection model. This limits the extension of our method to a larger set of images because in our initial analysis we observe that the DataComp face-bounding boxes provided in DataComp metadata often do not contain human faces.

\subsection{Cloudflare website categorization}
\label{appx:methods:cloudflare}

In \Cref{tab:cloudflare}, we show the merged category names or renamings based on the names returned by the Cloudflare Domain Intelligence API \citep{cloudflare}.

\begin{table}[ht]
\caption{A mapping of category names to the corresponding Cloudflare categories. All other category names follow Cloudflare naming \citep{cloudflare}. We follow categorization merging from \citet{ruth2022world} and extend to new or confusingly-named Cloudflare categories.}
\label{tab:cloudflare}
\begin{center}
\begin{tabular}{ll}
\hline
\textbf{Merged category name} & \textbf{Cloudflare category names} \\
\hline
Arts & Arts; Fine Art \\
Audio Streaming & Audio Streaming; Radio \\
Business & Business; Professional Networking \\
Chat \& Messaging & Chat; Instant Messengers; Messaging \\
Clothing \& Fashion & Clothing; Fashion; Lingerie \& Bikini; Swimsuits \\
Drugs \& Alcohol & Alcohol; Drugs; Tobacco \\
File Sharing & File Sharing; Photo Sharing \\
Government \& Politics & Government; Military; Politics, Advocacy, and Government-Related \\
Movies \& Home Video & Home Video/DVD; Movies \\
NSFW & Adult Themes; CIPA Filter; Militancy, Hate \& Extremism; Nudity; Pornography; Violence; Weapons \\
Science & Science; Space \& Astronomy \\
Stock Photos & News, Portal \& Search \\
Technology & APIs; Artificial Intelligence; Information Security; Information Technology; Technology \\
Video Streaming & P2P; Video Streaming \\
\\ \hline
\end{tabular}
\end{center}
\end{table}

\section{Additional results}
\label{appx:results}

Here we present additional results from our analysis in the manuscript.

\subsection{Detected languages}
\label{appx:results:language}

\begin{figure}[ht]
    \begin{center}
    \includegraphics[width=0.8\linewidth]{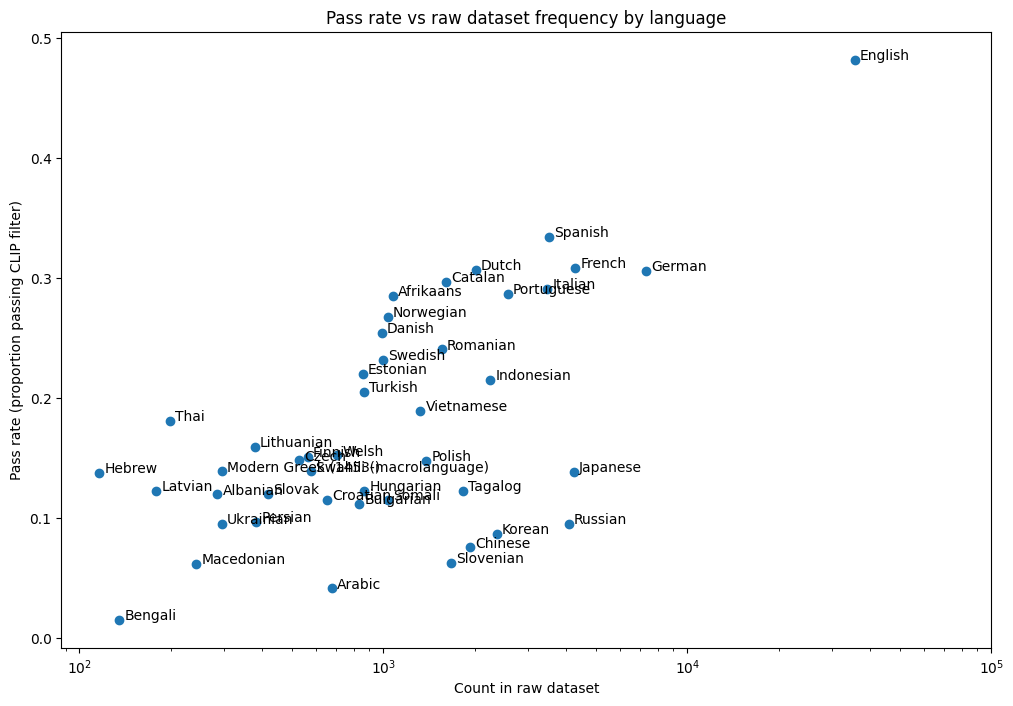}
    \end{center}
    \caption{Pass rate vs raw dataset frequency for a more comprehensive set of languages with at least 100 samples in the raw dataset. We observe a positive trend ($p < 0.001$) in which the more represented a language is in the raw dataset then the higher the pass rate.}
    \label{fig:language_more}
    \Description{A scatter plot of pass rate versus raw dataset frequency for languages with at least $100$ samples. Western languages have higher pass rates, and text detected as English comprise of $35,514$ samples (out of the $100,000$ samples) in the raw dataset with a pass rate of $48.2\%$. Most languages are in plotted in lower frequency magnitudes with pass rates ranging from $0$ to $0.5$.}
\end{figure}

\Cref{fig:language_more} plots the pass rate versus the raw dataset frequency for all detected languages that appear in at least $100$ samples (out of the $100,000$ randomly-chosen samples). Again, we find that the detected languages with the highest pass rates are Western languages, and a statistically-significant positive relationship between pass rate and raw dataset frequency. This illustrates the \textit{exclusion amplication} trend we find, where underrepresented languages are excluded at higher rates.

\subsection{Western bias on English data}
\label{appx:results:english}

\begin{figure}[ht]
    \centering
    \begin{subfigure}{0.45\linewidth}
         \includegraphics[width=\linewidth]{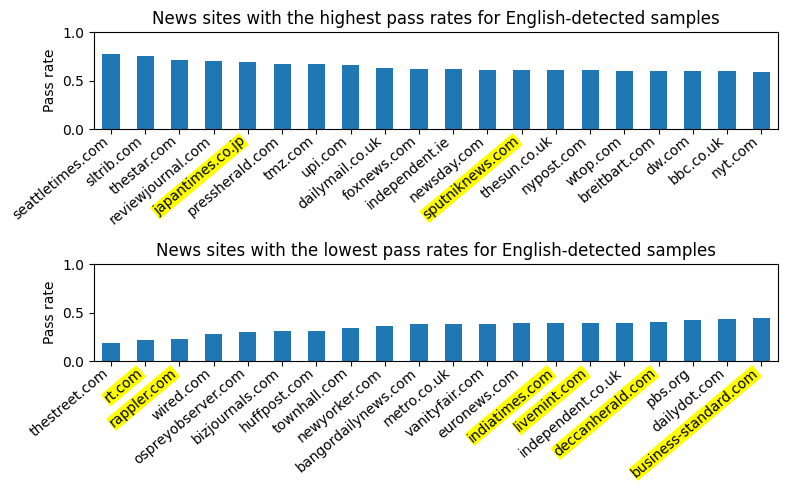}
        \caption{New sites with the highest pass rates (top) and lowest pass rates (bottom) for English-detected samples.}
    \label{fig:newsite_english}
    \end{subfigure}
    \hfill
    \begin{subfigure}{0.45\linewidth}
        \includegraphics[width=\linewidth]{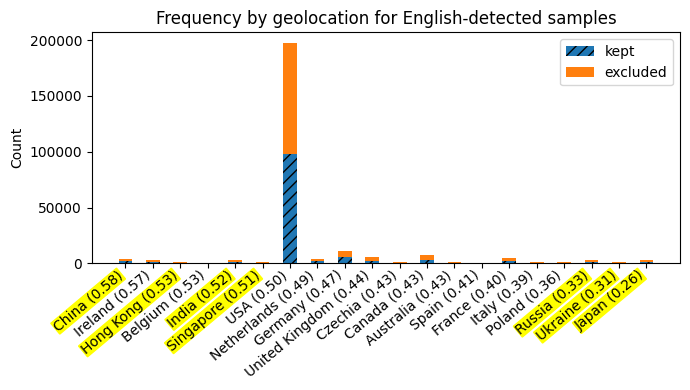}
        \caption{Frequency of English-detected data by country of IP address sorted by pass rates in descending order.}
        \label{fig:ip_addr_english}
    \end{subfigure}
    \hfill
    \begin{subfigure}{0.45\linewidth}
        \includegraphics[width=\linewidth]{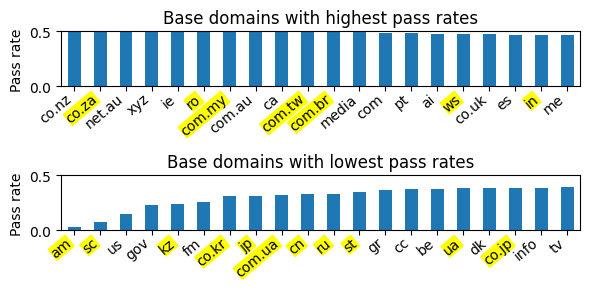}
        \caption{Pass rates for base domains with the highest (top) and lowest pass rates (bottom) for English-detected data.}
        \label{fig:base_domain_english}
    \end{subfigure}
    \caption{Pass rate by news site (a), IP address geolocation (b), and country domain (c) on English-detected data. Groups that correspond to non-Western regions are highlighted in yellow. We observe that some non-Western regions now have high pass rates, although data relating to many non-Western regions are included by the CLIP filter at low rates.}
    \label{fig:english}
    \Description{Bar graphs of pass rates by news site on the top left, IP address geolocation on the top right, and country domain at the bottom on English-detected data.}
\end{figure}

When examining filter discrepancies on data relating to different geographic regions, we isolate our analysis to English-detected data (via \texttt{langdetect} \citep{langdetect}). \Cref{fig:newsite_english}, \Cref{fig:ip_addr_english}, and \Cref{fig:base_domain_english} show the pass rates by news site, IP address geolocation, and country domain, respectively. We find that similar trends of Western bias hold, although much lower in magnitude and not across all dimensions of analysis.

\subsection{Examples of sexually-explicit text}
\label{appx:results:nsfw_examples}

\Cref{tab:nsfw} displays samples of sexually-explicit text that we manually find among the data from websites categorized as NSFW by Cloudflare. These samples are not caught by the NSFW filter, and some of them pass the CLIP filter as well.

\begin{table}[ht]
\caption{Examples of sexually-explicit text in CommonPool (i.e. passes initial NSFW filter). Names and locations are redacted.}
\label{tab:nsfw}
\begin{center}
\begin{tabular}{ll}
\hline
\textbf{Pass CLIP filter} & \textbf{Text} \\
\hline
Yes & **** Cleavage \\
Yes & **** caught by my spy cam taking a shower \\
Yes & ... you\_gonna\_get\_raped ... \\
No & Teen at gyno Thumbnail \\
No & Sexy **** teen girls \\
No & **** Nude Leaks
\\ \hline
\end{tabular}
\end{center}
\end{table}

\subsection{Internet Archive earliest-indexed date}
\label{appx:results:year}

\begin{figure}[ht]
    \centering
    \begin{subfigure}{0.45\linewidth}
        \includegraphics[width=\linewidth]{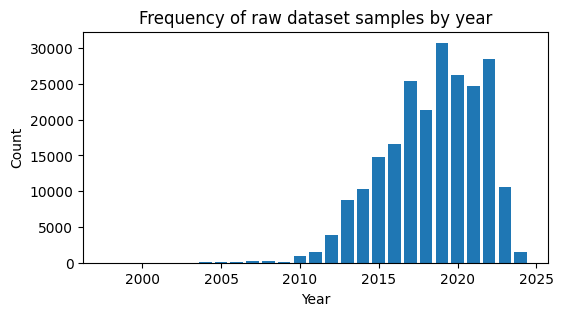}  
        \caption{Frequency by year in raw dataset. A majority of the samples examined come from the years 2019 - 2024.}
        \label{fig:year_raw_total}
    \end{subfigure}
    \hfill
    \begin{subfigure}{0.45\linewidth}
         \includegraphics[width=0.95\linewidth]{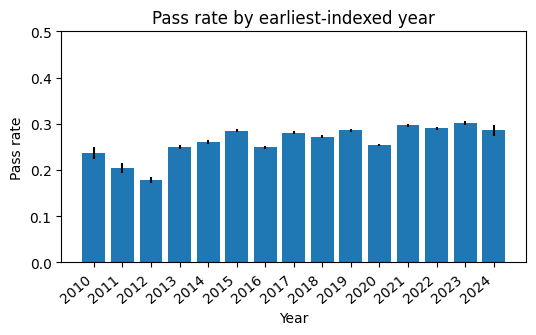}
        \caption{Pass rate by year for years with at least 500 associated samples. From 2010 - 2024 there is a slight positive trend in pass rate.}
        \label{fig:year_ratios}
    \end{subfigure}
    
    \caption{Raw dataset frequency (a) and pass rate (b) by earliest-indexed year in the Internet Archive. A higher pass rate represents a higher proportion included in the resulting CLIP-filtered dataset.}
    \label{fig:year}
    \Description{Bar graphs of raw dataset frequency on the left and pass rate on the right by earliest-indexed year in the Internet Archive. A higher pass rate represents a higher proportion included in the resulting CLIP-filtered dataset. The left bar graph shows the counts steadily increasing from $1000$ to $30000$ samples in years 2010 to 2022. The right bar graph shows the pass roughly increasing from $0.2$ to $0.3$ in years 2010 to 2024.}
\end{figure}

In \Cref{fig:year}, we demonstrate that most samples in the Internet Archive are first indexed in the last five years. Moreover, when we limit to years with at least 500 samples (i.e. 2010--2024), we observe a slight increasing trend in pass rates. More recent years have slightly higher pass rates than prior years and at the same time are more popular in the raw dataset.

\subsection{Occupation}
\label{appx:results:occupation}

\begin{figure}[ht]
   \begin{subfigure}{0.45\linewidth}
     \centering
     \includegraphics[width=\linewidth]{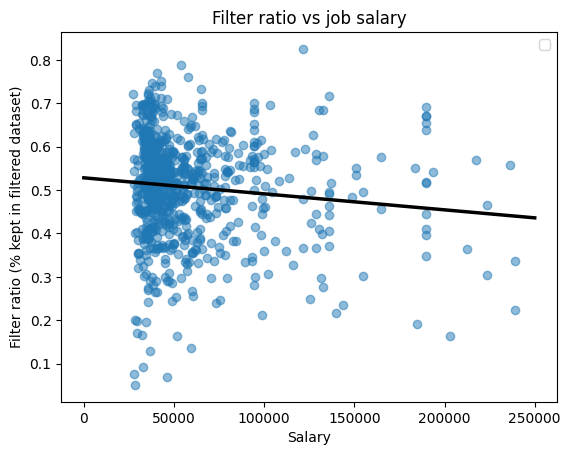}
     \caption{Pass rates vs job salary for O*NET job titles. $p = 0.001$}
     \label{fig:job_salary}
   \end{subfigure}
    \hfill
   \begin{subfigure}{0.45\linewidth}
     \centering
     \includegraphics[width=\linewidth]{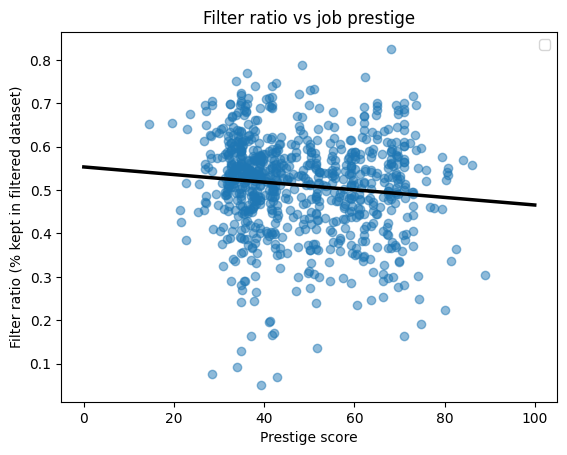}
     \caption{Pass rates vs job prestige for O*NET job titles. $p = 0.002$}
     \label{fig:job_prestige}
   \end{subfigure}
   \caption{Pass rates by occupations according to O*NET job titles \citep{onet} following analysis from \citet{aboutme}.}
   \Description{Two scatter plots of pass rate vs salary and pass rate vs prestige for O*NET job titles. There is a slight negative association with $p < 0.05$ for both. Job salaries range from $25000$ to $250000$ with high concentration on the lower end, prestige scores range from $15$ to $90$ spread consistently throughout, and pass rates range from $0$ to $0.9$ with high concentration in the middle.}
\end{figure}

We examine text samples in CommonPool that mention occupations. With the O*NET job titles, salary, and prestige \citep{onet}, we find in \Cref{fig:job_salary} and \Cref{fig:job_prestige} slight statistically-significant negative relationships. In other words, mentions of occupations with higher salary or higher prestige are more likely to be excluded. While this counters the findings from  obtained from \citet{aboutme}, these negative relationships are slight, however, so it is difficult to form a meaningful observation.

\subsection{Dialect}
\label{appx:results:dialect}

\begin{figure}[H]
    \begin{center}
    \includegraphics[width=0.5\linewidth]{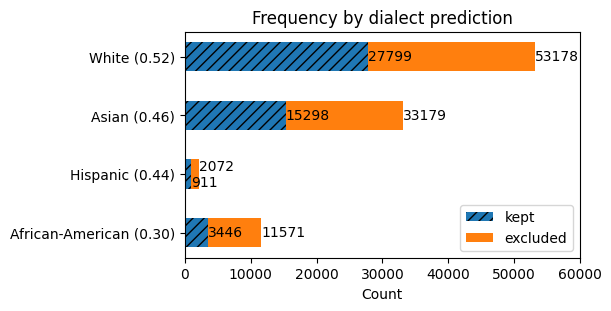}
    \end{center}
    \caption{Frequency by imputed dialect applied to a random subset of 100,000 samples detected with English text.}
    \label{fig:dialect}
    \Description{A bar graph of frequency by imputed dialect applied to a random subset of $100,000$ samples detected with English text. The White-imputed dialect has $53178$ samples total with a pass rate of $0.52$, the Asian-imputed dialect has $33179$ samples total with a pass rate of $0.46$, the Hispanic-imputed dialect has $2072$ samples total with a pass rate of $0.44$, and the African-American-imputed dialect has $11571$ samples total with a pass rate of $0.30$.}
\end{figure}

\Cref{fig:dialect} illustrates the frequency and pass rates by the dialect prediction model from \citet{blodgett2016demographic} on a random subset of $100,000$ samples detected as English by \texttt{langdetect} \citep{langdetect}. However, because the dialect prediction model was trained on Twitter data, upon manual examination we find that the dialect predictions are difficult to interpret with alt-text as input.

\section{Potential mitigations for dataset creators}
\label{sec:appx:interventions}

In this section, we list several actions that maintainers of web-scraped datasets like CommonPool may consider in order to mitigate the potential harms of NSFW, NCII, and CSAM content present in these datasets, as defined in \Cref{sec:discuss:consequences:nsfw}. These interventions may be components of the defense-in-depth approach discussed in \Cref{sec:discuss:recommendations:defense} and thus may be combined together, but cannot be considered complete solutions as they each contain tradeoffs. For each potential mitigation we detail their limitations and implications.

While we focus on the actions of dataset creators here, we also note that determining who is responsible for the presence of illegal or harmful data \citep{hutchinson2021towards, khan2022subjects} remains outside the scope of this work. Given that CommonPool sources from Common Crawl \citep{commoncrawl}, it is unclear how Common Crawl can address some of these NSFW concerns \citep{luccioni2021s}. In the case of dataset users (in other words, model developers), any update to the dataset may not impact models already trained on the original dataset. For instance, models trained on CSAM are not considered child sexual abuse material in US law \citep{laion_csam}, which explains why Midjourney and Stable Diffusion were not taken down when their training data was revealed to contain CSAM \citep{laion_takedown}. In addition, legal regulation of nonconsensual deepfake sexual imagery is still patchy, with many jurisdictions having no specific laws about deepfake NCII \citep{regulation_deepfake}. 

\subsection{Remove images from websites labeled as NSFW by Cloudflare}

In \Cref{sec:source:category:nsfw}, our audit finds that $1.03\%$ of CommonPool samples come from websites that Cloudflare labeled as NSFW (including pornography). These samples could easily be detected and removed from the original dataset, without checking or downloading image content. While simple to implement, we again caution that this method would not remove \textit{all} NSFW content and would not impact copies of the dataset already downloaded nor models already trained on portions of CommonPool. The NSFW-categorized domains corresponding to the subset of $400,000$ samples are available upon request.

\subsection{Run hash-based detection to known CSAM lists}
    
Dataset creators may follow perceptual hashing methods \citep{farid2021overview, thiel_identifying_nodate} to compare data samples to known hash sets of CSAM provided by existing organizations like National Center for Missing and Exploited Children (NCMEC) and Canadian Centre for Child Protection (C3P) \citep{ncmec, c3p}. Other APIs like Microsoft's PhotoDNA or Thorn's Safer tools can also be used to identify CSAM \citep{photodna, thorn}.

A lack of positive match here indicates that the dataset does not contain images that correspond to \textit{known} CSAM hashes, but does not reveal insight about \textit{unknown} instances of CSAM, much less instances of NCII or NSFW materials. If a sample is a positive match, there may be complications on its removal \citep{thiel_identifying_nodate}, as the dataset has likely been downloaded and copied many times. Removing CSAM then re-uploading the dataset could allow actors with old copies of the dataset to quickly identify CSAM \citep{laion_takedown}.

\subsection{Remove people from dataset}

Even if all instances of CSAM could be removed from datasets, the problematic ability of models to generate CSAM remains. Due to the capabilities of generative models to compose effectively \citep{okawa2024compositional}, prior investigation shows that bad actors may distort benign images of children into sexualized images or use de-aging techniques on NSFW-generated content, which would also circumvent prompt checks \citep{csam_companies, thorn_csam}. Removing all humans from training datasets would mitigate both CSAM in training data and generated imagery. Such an intervention would also mitigate some other deepfake harms, including deepfakes of political figures \citep{aoc}. Automated implementation on a large-scale dataset, however, would not guarantee complete removal, as existing person and face detection models have their own biases and error rates \citep{wilson2019predictive, rekogbias}.

\subsection{Restrict dataset license to research-only usage}

\citet{datacomp} release CommonPool as an index of image url-text pairs under a Creative Commons CC-BY-4.0 license \citep{ccby}, which allows for the commercial usage of CommonPool, including the deployment of text-to-image generative models trained on this dataset. To avoid potential harms of future deployed models \citep{thiel2023generative}, one avenue is to limit usage to research purposes only or adopting licenses with behavioral use conditions like the RAIL license \citep{rail}. This approach does not remedy the presence of harmful content, but at least prevents models trained on these datasets from being used at scale.

\subsection{Rebuild dataset from ground up}

In general, we find that the filtering approach, in which undesirable content is iteratively removed from a web dump, results in the potential for false negatives. As we show in \Cref{sec:source:category:nsfw}, a filtering mechanism may miss harmful content and thus include it in the final dataset. Another perspective to dataset collection is to rebuild the dataset from the ground up by adding only data known to be safe and ethically obtained. We point to the ante-hoc frameworks and recommendations from ethical dataset curation research \citep{andrews2024ethical, parrots, scheuerman2021datasets}. An example of this implementation is the inclusion of data from moderated sources like reputable news or stock photo sites. At the same time, however, sources may still include harmful content, and this method may not address copyright concerns \citep{copyright}.
    
\subsection{Retract the dataset}

Large-scale image datasets have been taken down in the past as a result of the discovery of CSAM or NCII -- for instance, LAION-5B \citep{laion_takedown} due to analysis by \citet{thiel_identifying_nodate} and Tiny Images \citep{tiny_takedown} based on the audit by \citet{birhane2021large}. While this may prevent future distribution of problematic content contained in the dataset, prior work shows that several previously-deprecated datasets have still been circulated via derivative copies and used in peer-reviewed research \citep{peng2021mitigating, luccioni2022framework}, 

We also note that the removal of open-access datasets would still allow the existence of similar proprietary datasets, which may have the same issues, to remain in usage without the ability for researchers to audit. In this manner, researchers would no longer have access to these datasets and therefore would not be able to study machine learning practices or models that may correspond to commercial approaches. We raise the question of what the goals of research on open datasets are, and when, if ever, the growing body of evidence of the fundamental flaws of web-scraped datasets will be enough.

\end{document}